\documentclass[amsmath,amssymb,aps,prd,twocolumn,showpacs,nolongbibliography]{revtex4-2}
\usepackage{graphicx}
\usepackage{dcolumn}
\usepackage{bm}
\usepackage{hyperref}
\usepackage{url}

\usepackage{breakurl}
\hypersetup{colorlinks=true}
\hypersetup{breaklinks=true}

\begin{document}

\title{ 
Fully general relativistic simulations of rapidly rotating quark stars: Oscillation modes and universal relations
}

\author{Kenneth Chen}
\email[]{kchen@link.cuhk.edu.hk}
\author{Lap-Ming Lin}
\email[]{lmlin@cuhk.edu.hk}
\affiliation{Department of Physics, The Chinese University of Hong Kong, Hong Kong, China}

\date{\today}

\begin{abstract}
Numerical simulation of strange quark stars (QSs) is challenging due to the strong density discontinuity at the stellar surface. In this paper, we report successful simulations of rapidly rotating QSs and study their oscillation modes in full general relativity. 
Building on top of the numerical relativity code \texttt{Einstein Toolkit}, we implement a positivity-preserving Riemann solver and a dustlike atmosphere to handle the density discontinuity at the surface. 
The robustness of our numerical method is demonstrated by performing stable evolutions of rotating QSs close to the Keplerian limit and extracting their oscillation modes. 
We focus on the quadrupolar $l=|m|=2$ $f$-mode and study whether they can still satisfy the universal relations recently proposed for rotating neutron stars (NSs).  
We find that two of the three proposed relations can still be satisfied by rotating QSs. 
For the remaining broken relation, we propose a new relation to unify the NS and QS data by invoking the dimensionless spin parameter $j$. 
The onsets of secular instabilities for rotating QSs are also studied by analyzing the $f$-mode frequencies. 
Same as the result found previously for NSs, we find that QSs become unstable to the Chandrasekhar-Friedman-Schutz instability when the angular velocity of the star $\Omega \approx 3.4 \sigma_0$ for sequences of constant central energy density, where $\sigma_0$ is the mode frequency of the corresponding nonrotating configurations.   
For the viscosity-driven instability, we find that QSs become unstable when $j\approx 0.881$ for both sequences of constant central energy density and constant baryon mass. Such a high value of $j$ cannot be achieved by realistic uniformly rotating NSs before reaching the Keplerian limit. 
The critical value for the ratio between the rotational kinetic energy and gravitational potential energy of rotating QSs for the onset of the instability, when considering sequences of constant baryon mass, is found to agree with an approximate value obtained for homogeneous incompressible bodies in general relativity to within 4\%. 
\end{abstract}


\maketitle

\section{Introduction}
\label{sec:intro}

\subsection{Quark stars} 

Do strange quark stars (QSs) exist in nature? The question remains unanswered since the hypothesis that strange quark matter composed of $u$-, $d$-, and $s$-quarks may be the ground state of baryonic matter was proposed as early as fifty years ago \cite{Itoh:1970,Bodmer:1971,Witten_1984}. 
If strange quark matter is only metastable, then hybrid stars consisting of quark matter cores surrounded by nuclear matter in the envelope may also exist (e.g., \cite{Alford:2005,Alford2013}). 
More recently, the possibility that quark matter containing only $u$- and $d$-quarks is the true ground state of baryonic matter for a baryon number larger than 300 has also been considered \cite{Holdom:2018}.

While there is still no evidence for their existence, QSs have been proposed to explain some compact-object observations in the past \cite{Weber:2005}. 
More recently, the low-mass ($< 1 M_\odot$) central object of the supernova remnant HESS J1731-347 is suggested to be a QS \cite{Doroshenko:2022,DiClemente:2022,Horvath:2023}, as it is not possible to form such a low mass neutron star (NS) by conventional core-collapse supernova \cite{Suwa:2018}. 
In the era of gravitational wave (GW) astronomy, the event GW190814 \cite{GW190814} has also been suggested to be a black hole-QS system \cite{Bombaci:2021}. 
Constraints on the equation of state (EOS) of quark matter
have also been considered by assuming that the events GW170817 \cite{GW170817} and GW190425 \cite{GW190425} were due to merging QSs instead of NSs \cite{Miao:2021}.    
While the observed kilonova signal associated with GW170817 suggests that the event was due to a binary NS merger, this event by itself does not rule out the existence of QSs since NSs and QSs could coexist according to the two-families scenario \cite{Drago:2016,DePietri:2019}. 
Better constraints on the properties of quark matter or even direct evidence for QSs might be possible as more GW events are expected to be observed in the coming decade.

Numerical relativity simulations are indispensable tools for studying the GWs emitted from strongly dynamical spacetimes, such as the mergers of binary compact objects. 
While hydrodynamic simulations of NSs in full general relativity are performed routinely nowadays by different research groups (see \cite{Baiotti2017,Duez:2019,Kyutoku:2021} for recent reviews), only a few relativistic simulations have been obtained for QSs. 
The first binary QS simulation was done in 2009 \cite{Bauswein:2009} using the smooth particle hydrodynamics method and the conformally-flat approximation in general relativity. 
Fully general relativistic simulations of single and binary QSs \cite{Zhenyu2021,Enping2021,Enping2022} became available only in the past two years. 

In this paper, we add a contribution to this line of research by demonstrating our ability to evolve
rapidly rotating QSs and study their oscillation modes. 
Our simulations were performed using the publicly available code \texttt{Einstein Toolkit} \cite{ETpaper,ET_Turing,ETweb}, with our own implementation of a positivity-preserving Riemann solver and a dustlike EOS for the ``atmosphere."
Apart from the fact that the study of oscillation modes of compact stars is important in its own right (see below), the demonstration of stable evolutions of rapidly rotating QSs would be an important milestone for us to achieve before attempting generic nonlinear dynamical situations, such as the gravitational collapse of a rapidly rotating unstable QS.  

The challenge to evolve a bare QS (without a thin nuclear matter crust), as described by the standard MIT bag-model EOS in a hydrodynamic simulation, is due to its sharp high-density surface where the pressure vanishes. 
The high-density surface is directly in contact with the numerical atmosphere, which is introduced to fill up the vacuum space with the purpose of stabilizing traditional grid-based hydrodynamic simulations. 
The low-density atmosphere is considered to have a negligible impact on the dynamics of compact stars when the evolution time is relatively short and comparable to the dynamical timescale, such as in the case of binary inspiral and merger; however, its small effects, if not properly handled, would accumulate and eventually kill a long-time simulation of a single stable star. 

The large contact discontinuity at the QS surface due to the density can be regarded as a special case of shock waves. 
In the context of shock-capturing hydrodynamic schemes, it is well known that low-order Godunov-type schemes~\cite{godunov:hal-01620642,HartenLaxLeer1983} that are strongly dissipative will smear the shock, while high-order schemes will usually introduce spurious oscillations that result in the erroneous reconstruction of density near the surface. 
The error introduced in NS modeling is typically small but serious for QSs and can cause significant violations of mass conservation.

It is essential to preserve the positivity of density (and pressure) for a QS near its surface.
A fine balance could be achieved by combining the high-resolution shock-capturing methods together with the so-called positivity-preserving (PP) Riemann solver, which was first introduced into the numerical relativity community by Radice, Rezzolla, and Galeazzi in~\cite{Radice:2013xpa}. 
The main idea is that one could always build a finite-volume PP scheme by integrating a high-order solver with a first-order one under a more restrictive Courant-Friedichs-Lewy (CFL) condition as shown by Shu {\it et al.}~\cite{Perthame1996,ZHANG20108918,ZHANG20111238,HU2013169}, since first-order Godunov-type schemes are known to have the PP property~\cite{EINFELDT1991273}.
The PP scheme was originally designed to treat the low-density atmosphere of NSs. Better mass conservation and sharper surface density profiles were obtained.  
In this study, we applied the idea to QSs and achieved similar improvements, with a dustlike EOS designed for the atmosphere. 
As required by the continuity conditions, the atmospheric density may no longer be small but can be in the same order as the surface density. 
When we model the atmosphere by nearly pressureless dust particles, large truncation errors on the densities will not cause noticeable disturbance on pressure profiles. 
Our strategy to handle the sharp QS surface is different from those employed in recent simulations of QSs. In~\cite{Enping2021,Enping2022}, Zhou {\it et al.} modified the primitive-variable recovery 
procedure together with an addition of a thermal component to the cold MIT bag model EOS. 
On the other hand, Zhu and Rezzolla \cite{Zhenyu2021} introduced a thin crust described by a polytropic EOS at the QS surface. 

\subsection{Oscillations of compact stars}
Pulsations of compact stars are potential sources of GWs, and their detection can provide important information for the uncertain properties of supranuclear EOS inside a traditional NS. 
The detected signals may even provide evidence for the existence of deconfined quark matter, which could exist in the core of a hybrid star model or in the form of a pure strange QS~\cite{Witten_1984, Alford2013}.

The successful detection of GWs from merger events by the Laser Interferometer Gravitational Wave Observatory (LIGO)-Virgo Scientific Collaboration~\cite{virgo,LIGO,PhysRevLett.116.061102} has opened a new era of observational astronomy. Advanced LIGO, Virgo, Kamioka Gravitational Wave Detector, and the next-generation detectors such as the Einstein Telescope~\cite{EinsteinTelescope} would have sufficient sensitivities in the high-frequency band ($\sim {\rm kHz}$) to probe the GWs emitted from pulsating compact stars. 

The most important oscillation modes that would have interests in GW astronomy are the quadrupolar ($l=2$) fundamental $f$-mode, the first few overtones of the pressure $p$-modes, rotational $r$-mode, and maybe the first spacetime $w$-modes ~\cite{Kokkotas1999, Andersson2011}. 
The $f$-mode is particularly relevant to the GW signals emitted from isolated and binary NS systems. 
On the one hand, the $f$-mode is expected to contribute strongly to the GWs emitted from a proto-NS~\cite{Radice_2019,Morozova_2018}. 
For a binary NS system, the dynamical tidal effects due to the coupling between the excited $f$-mode and tidal fields during the late inspiral phase are important to the dynamics and emitted GW from the system \cite{Hinderer_2016,Steinhoff2016}. 

While the oscillation modes of nonrotating compact stars can be formulated and computed as eigenvalue problems (see \cite{Kokkotas1999} for a review), the situation for rapidly rotating stars is more complicated as the effects of rotation cannot be treated perturbatively. 
A standard approach to studying the oscillation modes of a rapidly rotating NS in general relativity is to suitably perturb the star and follow its subsequent evolution using a hydrodynamic code, and the mode frequencies are then identified from the Fourier spectra of the fluid variables. 
Due to the complexity of general relativity, such simulations are usually performed under the Cowling approximations \cite{Font:2001,Stergioulas2004,Gaertig:2008,DonevaGaertig_2013} and the conformally flat assumptions \cite{Dimmelmeier2006,Ng_2021}. 
An exception is the work by Zink {\it et al.} \cite{PhysRevD.81.084055} in 2010 which investigated the $f$-modes of uniformly rotating polytropic stars using a nonlinear hydrodynamics code in full general relativity. More recently, Kr\"uger and Kokkotas (hereafter KK) in~\cite{Kokkotas2020,PhysRevD.102.064026} studied the $f$-modes of rapidly rotating NSs with realistic EOSs taking into account the spacetime dynamics in a linearized theory. 
In this work, we shall study the oscillation modes of rapidly rotating QSs in full general relativity for the first time.  

Focusing on the quadrupolar ($l=2$) $f$-mode, the three $m=0,\pm2$ modes are degenerate for a nonrotating spherical star, where $m$ is the azimuthal quantum number. 
They will split when the star rotates, similar to the Zeeman splitting in quantum mechanics, 
though the splitting does not increase linearly with the rotation rate due to the high nonlinearity of the system. The two nonaxisymmetric ($m \neq 0$) modes are usually called bar modes, and they are subject to various instabilities \cite{Andersson_2003,Friedman:2013}. 
In this work, we shall determine the onsets of secular instabilities of rapidly rotating QSs driven by GW and viscosity dissipations. 

For the GW-driven Chandrasekhar-Friedman-Schutz (CFS) instability \cite{Chandrasekhar1970,FriedmanSchutz1978}, the onset occurs at a neutral point, where the counterrotating $m=2$ mode frequency $\sigma_i$ observed in the inertial frame passes through zero (i.e., $\sigma_i=0$).  
In \cite{Kokkotas2020}, KK found that the onset of CFS instability for a rotating NS occurs when the angular velocity $\Omega$ of the star $\Omega \approx 3.4 \sigma_0$ for sequences of constant central energy density, where $\sigma_0$ is the $f$-mode frequency of the corresponding nonrotating model. 
The conclusion is approximately insensitive to the chosen EOS models in their study. 
We shall see in our work that rapidly rotating QSs also satisfy this result as well. 

The bar modes are also subject to another type of instability which is driven by viscosity. 
The instability sets in when the $m=-2$ corotating mode frequency $\sigma_c$ in the rotating frame passes through zero (i.e., $\sigma_c=0$). 
The Newtonian analysis of the onset of this instability was studied by Chandrasekhar 
\cite{CHANDRASEKHAR_1969} for a sequence of uniformly rotating uniform-density Maclaurin spheroids. 
It was found that a new sequence of triaxial Jacobi ellipsoids branches off the Maclaurin sequence when the Newtonian ratio between the rotational kinetic energy $T$ and gravitational potential energy $|W|$ reaches the critical value $(T/|W|)_{\rm crit,Newt} = 0.1375$. 
Above this critical value, the Maclaurin spheroids are subjected to the viscosity-driven instability and migrate towards the Jacobi sequence by dissipating energy while conserving angular momentum.   
A Jacobi ellipsoid is particularly relevant to GW astrophysics, as its time-varying mass quadrupole moment will continuously emit GW radiation.

In \cite{PhysRevD.66.044021}, it is found that general relativity weakens the Jacobi-like bar-mode instability. Furthermore, a stiff EOS with an adiabatic index as large as $2.5$ is required for a $1.4 M_\odot$ polytropic star to become unstable for $\Omega$ lower than the Keplerian limit~\cite{Bonazzola1998}.  
The onset of the instability is thus expected to be difficult to achieve (if not impossible) by realistic rotating NSs. 
On the other hand, rotating QSs would be the most promising candidates to achieve the instability as they can generally support higher rotation rates \cite{Gourgoulhon_1999,Lo_2011} and are stiff enough to be approximated well by incompressible models \cite{Sham_2015}. 
 
The viscosity-driven instability of rotating QSs was already studied more than twenty years ago \cite{Gourgoulhon_1999,Gondek2000,Gondek2003}, and the instability onset was found to occur generally before the Keplerian limit.  
However, these studies were not based on the analysis of the oscillation modes, but by perturbing the stellar configuration during the iteration steps of the calculation of an axisymmetric equilibrium rotating star. 
If the perturbation grows during the iteration, then the star is declared to be unstable.  
In this work, we study for the first time the onset of the viscosity-driven instability by observing how the corotating mode frequency $\sigma_c$ in the rotating frame passes through zero as the rotation rates of sequences of QSs approach the Keplerian limit. 
It should be noted, however, that there is no physical viscosity in our simulations as all stars are modeled by perfect fluids. While the oscillation modes were identified in our three-dimensional simulations, the spontaneous breaking of axisymmetry due to the instability was not observed in the dynamical timescale.

\subsection{Universal relations} 
In the last decade, the discoveries of various approximate EOS-insensitive universal relations of compact stars (see \cite{Yagi2017,Doneva:2018} for reviews) are not only of theoretical interest but also of importance in astrophysical applications, such as measuring masses and radii with x-ray pulse profile modeling~\cite{Psaltis_2014}, analyzing GW signals to constrain the maximum mass of NSs~\cite{Rezzolla_2018}, and reducing the number of parameters in theoretical gravitational waveform models for binary NS inspirals \cite{Lackey:2019, Schmidt:2019,Barkett:2020,Pnigouras:2021}.
 
In contrast to the mass-radius relations of compact stars, universal relations connecting different physical quantities are generally insensitive to EOS models to about 1\% level.  
Many of the investigations done on universal relations only focus on traditional NSs, though it is known that bare QSs also satisfy some of the relations established by NSs 
\cite{Lau2010,Yagi2013,Yagi:2013b,Chan2014}. 
Besides searching for new universal relations, which may provide astrophysical applications, it is also interesting to test existing universal relations against different physics inputs such as thermal effects relevant to hot newborn NSs \cite{Martinon:2014,Marques:2017} or superfluid dynamics for cold NSs \cite{Yeung:2021}. 

Attempts to find universal relations for the oscillation modes of compact stars dated back to the seminal work of Andersson and Kokkotas \cite{AnderssonKokkotas1998} more than twenty years ago, which was then followed by Benhar {\it et al.} \cite{Benhar:2004} and Tsui and Leung \cite{Tsui:2005prl,Tsui:2005}.   
While these earlier universal relations depend weakly on the EOS models to a certain accuracy, they are not as robust as those discovered later. 
The $f$-mode is now known to connect to the moment of inertia \cite{Lau2010} and tidal deformability \cite{Chan2014} by robust universal relations which are insensitive to the EOS models to within about 1\% level, when the relevant physical quantities are suitably scaled (see, e.g., \cite{Sotani:2021,Lioutas:2021,Zhao:2022,Kuan:2022} for recent work).
However, these studies were based on nonrotating NSs and QSs only. 
Recently, KK \cite{Kokkotas2020} found three universal relations for the bar modes of rapidly rotating NSs using their newly developed code that takes into account spacetime dynamics in a linearized theory \cite{PhysRevD.102.064026}. 
In this paper, we shall study whether their universal relations can also be applied to rapidly rotating QSs. We find that two of their relations can still be satisfied by bare QSs very well, but one of them is broken quite significantly already at moderate rotation rates. 
In addition to the $f$-mode, we also study the first $p$-mode of rotating QSs. For the class of QS models studied in this paper, we report fitting relations for the $p$-mode frequencies of both nonrotating and rotating stars.


The plan of the paper is as follows. 
In Sec.~\ref{sec:formulation}, we discuss the formulation and numerical methods employed in this work. 
Section~\ref{sec:TestAndResult} presents the numerical results, including tests that were performed to validate
our simulations. Finally, we conclude the paper in Sec.~\ref{sec:conclude}. 
Unless otherwise noted, we adopt the unit convention $c=G=M_\odot=1$, where $c$ is the speed of light, $G$ is the gravitational constant, and $M_\odot$ is the solar mass.

\section{Formulation and numerical methods}
\label{sec:formulation}
Our simulations were performed using the publicly available code \texttt{Einstein Toolkit} which is built on top of the \texttt{CACTUS} computational infrastructure~\cite{cactus2003,Cactusweb}.
The spacetime is evolved using the standard CCZ4 formulation of the Einstein equations~\cite{Alic2012,Alic2013} implemented in the thorn code \texttt{McLachlan}~\cite{Brown:2008sb,mlweb} of \texttt{Einstein Toolkit}. 
We choose the parameters of the CCZ4 formulation to be $\kappa_2 = 0$ and $\kappa_3 = 0.5$ in our simulations. As for $\kappa_1$, we typically choose it to be $0.05$. 
Although its optimal value can vary for different models, physical results are insensitive to these choices as long as the constraint violation does not grow and invalidate the simulations. 
The general relativistic hydrodynamics equations are solved using the thorn code
\texttt{GRHydro}~\cite{Baiotti:2004wn,M_sta_2013}. The mesh-refinement driver \texttt{CARPET}~\cite{Schnetter_2004,CarpetCode:web} is employed to provide an adaptive mesh refinement approach to increase resolution.  
The standard gauge conditions ``1$+$log" slicing~\cite{PhysRevLett.75.600} and Gamma-driver shift condition~\cite{PhysRevD.73.124011} are adopted, where the damping coefficient which is introduced to avoid strong oscillations in the shift is chosen to be $1/M$ (with $M$ being the gravitational mass).
Furthermore, a numerical dissipation of the Kreiss-Oliger type~\cite{kreiss1973methods} is introduced for spacetime variables and gauge quantities following the suggestion of~\cite{BaiottiRezzolla2006}.
The formulation and numerical setup used in our simulations are quite standard choices for general relativistic hydrodynamic modelings, such as in the cases of binary neutron star mergers. 
In order to simulate rapidly rotating QSs for sufficient duration in our study, we implemented a positivity-preserving Riemann solver to the \texttt{GRhydro} thorn which will be discussed below. 

\subsection{Positivity preserving Riemann solver}
\label{sec:ppsolver}

The fluid-vacuum interface at the surface of a star in hydrodynamics modeling is subject to perturbations mainly due to truncation errors.  
These perturbations could be significant when the surface has a nonzero finite density, as in the situation for a bare QS. 
This is particularly the case for simulations using Cartesian coordinates, where the grid points do not match well with the smooth stellar surface. 
When a free boundary condition is used, 
the freely evolved vacuum would quickly encounter numerical problems as it may lead to nonphysical negative densities. 
The problem is tackled, in general, by introducing an artificial atmosphere with a floor density $\rho_f$ in hydrodynamic simulations.  

It is therefore necessary and desirable to preserve the positivity of certain hydrodynamical variables, essentially the density and the pressure, in a free evolution scheme. 
For the Newtonian Euler equation, it has been shown that both the density and the pressure are guaranteed to be positive by a well-designed limiter when no source terms are present~\cite{HU2013169}. 
In~\cite{ZHANG20111238}, four types of source terms were tested for the discontinuous Galerkin schemes.
However, in relativistic hydrodynamics, a rigorous strategy is still lacking for a generic EOS. 
Here we discuss the PP Riemann solver introduced in~\cite{Radice:2013xpa,HU2013169}, which we implemented and proved to provide stable evolutions of rapidly rotating QSs in the study. 


Let us first give an outline of the conservative form of the relativistic hydrodynamics equations to define the variables for further discussion.
The standard $3+1$ Arnowitt-Deser-Misner form~\cite{Arnowitt1962} of spacetime metric is given by \begin{equation*}
    ds^2 = g_{\mu\nu}dx^\mu dx^\nu = (-\alpha^2+\beta_i\beta^i) dt^2 + 2\beta_i dt dx^i + \gamma_{ij} dx^i dx^j,
\end{equation*} 
where $g_{\mu\nu}$, $\alpha$, $\beta^i$, and $\gamma_{ij}$ are the spacetime 4-metric, lapse function, shift vector, and spatial 3-metric respectively. 
The energy-momentum tensor of the matter inside the star is assumed to take the perfect fluid form~\cite{M_sta_2013}
\begin{equation}
    T_{\mu\nu} = \rho h u_\mu u_\nu + P g_{\mu\nu}, 
\end{equation} where $\rho$ is the rest-mass density, $u^\mu$ is the fluid four-velocity, $P$ is the pressure, 
$h=1+\epsilon+P/\rho$ is the specific enthalpy, and $\epsilon$ is the specific internal energy.  
The equations of motion for the fluid are the conservation law of baryon number, i.e., Eq.~(\ref{eq:cons1}), and the conservation of energy and momentum, i.e., Eq.~(\ref{eq:cons2}),
\begin{subequations}
\begin{eqnarray}
\label{eq:cons1}
\nabla_\mu(\rho u^\mu)&=&0  ,  \\
\label{eq:cons2}
 \nabla_\mu T^{\mu \nu} &=& 0  ,
\end{eqnarray}
\end{subequations}
which are solved by the Valencia formulation~\cite{PhysRevD.43.3794,Banyuls_1997} in the conservative form, \begin{equation}
    \frac{\partial \mathbf{U}}{\partial t} + \frac{\partial \mathbf{F}^i}{\partial x^i} = \mathbf{S},
    \label{eq:hydro_conserv_form}
\end{equation}
with the conserved variables 
\begin{equation}
    \mathbf{U} = \left[ D,S_j,\tau \right] = \sqrt{\gamma} \left[ \rho W,\rho h W^2 v_j, \rho h W^2- P -\rho W \right] ,
\end{equation} 
where $\gamma$ is the determinant of $\gamma_{ij}$, and the three-velocity is $v^i = ({u^i}/{u^t} + {\beta^i})/\alpha$; $W=({1-v^iv_i})^{-1/2}$ is the Lorentz factor. For three-vectors like $v^i$ and $\beta^i$, their indices are raised and lowered by the 3-metric, e.g., $v_i = \gamma_{ij}v^j$.
The fluxes are \begin{equation}
    \mathbf{F}^i = \alpha [D\tilde{v}^i,S_j\tilde{v}^i + \sqrt{\gamma}P\delta^i_j,\tau\tilde{v}^i+\sqrt{\gamma}P v^i],
\end{equation}
and the source functions are 
\begin{equation}
\label{eq:source}
    \mathbf{S} = \alpha\sqrt{\gamma} [0,T^{\mu\nu}(\partial_\mu g_{\nu j}-\Gamma^\lambda_{\mu\nu} g_{\lambda j}),\alpha(T^{\mu 0}\partial_\mu \ln\alpha -T^{\mu\nu} \Gamma^0_{\mu\nu})] ,
\end{equation}
where $\tilde{v}^i=v^i - \beta^i/\alpha$ and $\Gamma^\lambda_{\mu\nu}$ are the 4-Christoffel symbols.

To illustrate the idea of the PP scheme, we first consider a source-free scalar conservation law in one dimension \cite{Radice:2013xpa}
\begin{equation}
    \frac{\partial u}{\partial t} + \frac{\partial f(u)}{\partial x} = 0.
\end{equation}
A theory~\cite{sigal2001,Hesthaven2008} states that a high-order temporal integration scheme such as the 
Runge-Kutta scheme which is a convex combination of forward Euler steps will maintain the total variation diminishing (TVD) property and the positivity of $u$, provided this is true for the first-order forward Euler method. Methods of this kind are known as strong stability-preserving or TVD methods. 
Considering a discretization scheme using the forward Euler method 
\begin{equation}
    \frac{u_i^{n+1}-u_i^n}{\Delta t} = \frac{f_{i-1/2}-f_{i+1/2}}{\Delta x},
\end{equation}
we can arrange it in the form
\begin{equation}
    u_i^{n+1} = \frac{1}{2}(u_i^+ + u_i^-) , 
\end{equation}
where 
\begin{subequations}    \label{eq:upm}
\begin{eqnarray}
u_i^+ &=&  u_i^n+2\frac{\Delta t}{\Delta x}f_{i-1/2}, \\
u_i^- &=&  u_i^n-2\frac{\Delta t}{\Delta x}f_{i+1/2}.
\end{eqnarray}
\end{subequations}
Sufficiently, when both $u_i^+$ and $u_i^-$ are positive, so will be $u_i^{n+1}$.
It is proven that positivity is guaranteed for the first-order Lax-Friedrichs (LF) flux~\cite{Perthame1996} with a more restrictive CFL-like condition ${\Delta t}/{\Delta x}\le 1/2c$, where $c$ is the largest speed of sound~\cite{HU2013169}. However, this low-order scheme is too dissipative to capture features of shocks.
The principle of the PP solver is to combine it with a high-order (HO) scheme for optimization,
\begin{equation}
\label{eq:ppsolver}
    f_{i+1/2}^{\rm{PP}} = \alpha f_{i+1/2}^{\rm{HO}} + (1-\alpha) f_{i+1/2}^{\rm{LF}},
\end{equation}
where $f_{i+1/2}^{\rm{HO}}$ is the HO flux, $f_{i+1/2}^{\rm{LF}}$ is the LF flux, and $\alpha\in[0,1]$ is an undetermined coefficient. 
In our simulations, we selected the Marquina solver~\cite{DONAT199642,Aloy_1999} as our HO flux. 
When $\alpha=1$, the PP scheme fully restores to the powerful HO flux, which is applied for the bulk of a star. On the other hand, around the fluid-atmosphere interface at the stellar surface, the PP scheme then searches for the optimal value of $\alpha$, compromising accuracy for positivity. 

The first component of Eq.~(\ref{eq:hydro_conserv_form}), i.e., the continuity equation, is source-free 
[see Eq.~(\ref{eq:source})] and the PP scheme is applicable.
Its conserved variable is $D=\sqrt{\gamma}W \rho$, where $\gamma$ and $W$ are both definitely positive. Ensuring the positivity of $D$ will serve our purpose of ensuring the positivity of $\rho$. 
However, the pressure-related term, the conserved energy density $\tau$, has a complex source term in Eq.~(\ref{eq:source}). 
While the authors of~\cite{Radice:2013xpa} suggested enforcing a floor value on $\tau$, empirically we found it adequate to apply the PP limiter also on $\tau$.
In our three-dimensional Cartesian-grid case, the CFL-like condition becomes ${\Delta t}/{\Delta x}\le 1/6c$, and the PP flux $f_{i+1/2}^{\rm PP}$ is calculated component-by-component.
This condition, which requires each interface to be non-negative, is too restrictive since only the positivity of their sum is really demanded. 
It could tolerate a small negative contribution (if any) from the source term. 

Let the conserved variable $u$ represent either $D$ or $\tau$.
The value of $\alpha$ is determined as follows. 
If $u_{i+1}^+(f_{i+1/2}^{\rm{HO}})$ is positive, then $\alpha(u_{i+1}^+)=1$, meaning 
that the original HO flux is used. Otherwise, 
\begin{equation}
    \alpha(u_{i+1}^+) = \frac{u_{i+1}^+({f_{i+1/2}^{\rm{LF}}})}{u_{i+1}^+(f_{i+1/2}^{\rm{LF}})-u_{i+1}^+(f_{i+1/2}^{\rm{HO}})}  , 
\end{equation} 
and similarly for $\alpha(u_i^-)$.
The PP property of the LF scheme ensures that a solution $\alpha\ge 0$ always exists. We then have
\begin{equation}
    \alpha(u) = \min\left(\alpha(u_{i+1}^+),\alpha(u_i^-)\right).
\end{equation} 
This determines the PP flux $f_{i+1/2}^{\rm PP}$ of one component.
As there are five components in the flux $\mathbf{F}^i$, different values of $\alpha$ could be applied to different components. Empirically, we found that using the smaller value between $\alpha(D)$ and $\alpha(\tau)$ on all five components worked well.
Practically, a smaller Courant factor or a more conservative choice of $\alpha$ can always be chosen if there is an intolerably large violation of mass conservation, which indicates a poor preservation of positivity.
The implementation of the PP solver allows us to set the floor density of the atmosphere to be $\rho_f=10^{-18}$ which is about $10^{-15}$ of the typical central density. 
In theory, the PP scheme allows the atmosphere to evolve freely to be as small as the round-off precision. In our typical simulations, the violation of the total mass conservation near $t=2000\approx 9.85$ ms for a Courant factor $0.16$ is of $\mathcal{O}(0.1\%)$. More numerical tests are presented in Sec.~\ref{sec:TestAndResult}.

\subsection{Equation of state}
\subsubsection{For QSs and NSs}
\begin{table}[t]
\begin{ruledtabular}
\caption{\label{table:eos1} Parameters of the four quark-matter MIT bag EOSs indicated by MIT1, MIT2, MIT3, and MIT4.}
\begin{tabular}{ccccc}
&MIT1&MIT2&MIT3&MIT4 \\
\hline
$c_{ss}$&1/3&1&2/3&1/2\\
$B/B_{60}$&1&3&3/2&3/2\\
\end{tabular}
\end{ruledtabular}
\begin{ruledtabular}
\caption{\label{table:eos2} Parameters of the piecewise polytropic representation of the nuclear-matter SFHo EOS. The parameters $(\rho_i, K_i, \Gamma_i, a_i)$ are expressed in the code units where $G=c=M_\odot=1$. }
\begin{tabular}{ccccc}
$i$ & $\rho_i$ & $K_i$ & $\Gamma_i$ & $a_i$ \\
\hline
0 & 0.0 & $6.0073\times10^{-2}$ & 1.2524 & 0.0 \\
1 & $8.4981\times10^{-7}$  & $7.4003\times10^{-6}$ & 0.6084 & $1.1491\times10^{-2}$ \\
2 & $4.2591\times10^{-6}$  & $5.4052\times10^{-3}$ & 1.1416 & $2.4695\times10^{-3}$ \\ 
3 & $5.3619\times10^{-5}$  & $2.3751\times10^{2}\ \ $  & 2.2288 & $1.0860\times10^{-2}$ \\ 
4 & $4.2591\times10^{-4}$  & $3.6338\times10^{4}\ \ $  & 2.8769 & $1.5677\times10^{-2}$ \\
\end{tabular} 
\end{ruledtabular}
\end{table}

As we are interested in extracting the oscillation modes of QSs excited by small perturbations, 
the thermal effects like shock heating will play a negligible role in the simulations. 
We thus assume the stars are described by zero-temperature EOS models.  
In order to model bare QSs, we parametrize the linear approximation of the MIT bag model EOS~\cite{Zdunik2000strange} by the square of the speed of sound $c_{ss}$ and the bag constant $B$ as 
\begin{equation}
\label{eq:eos}
 P = c_{ss} e - (1+c_{ss}) B  , 
\end{equation}
where $P$ is the pressure for a given energy density $e$. 
It will be convenient to further parametrize it by the ratio $\kappa \equiv \rho/\rho_S$ between the rest-mass density $\rho$ and the surface density at zero pressure $\rho_S=(1+c_{ss})B/{c_{ss}}$.
The full EOS is then given by  
\begin{subequations}
\label{mit:eos}
\begin{eqnarray}
    \rho&=& \rho_S \kappa,\\
    P&=& B(\kappa^{1+c_{ss}}-1),\\ 
    e&=& B \left( \frac{\kappa^{1+c_{ss}}}{c_{ss}}+1
    \right),
\end{eqnarray}
\end{subequations}
and $\kappa$ is closely related to the enthalpy $h$ through the relation 
\begin{equation}
    h=\frac{e+P}{\rho}=\kappa^{c_{ss}} .
\end{equation} 
In addition to a conventional choice of $c_{ss}=1/3$ and $B = B_{60} \equiv 60\  \rm{MeV/fm}^{3}$, hereafter denoted by ``MIT1," we also include the following models:
``MIT2" for $c_{ss}=1$ and $B/B_{60}=3$, ``MIT3" for $c_{ss}=2/3$ and $B/B_{60}=3/2$ and
``MIT4" for $c_{ss}=1/2$ and $B/B_{60}=3/2$, to cover a range of parameter space for QSs as we shall also explore the robustness of some EOS-insensitive universal relations. 

We also include a nuclear-matter EOS model for NSs as a benchmark for comparison. 
Instead of using the original tabular EOS data, we use a piecewise polytropic model~\cite{Read_2009} to represent analytically the SFHo EOS~\cite{Hempel2010,SFH2013}, which was not included in the study of the universal relations of rapidly rotating NSs by KK~\cite{Kokkotas2020}. 
In particular, we construct a five-piece model so that the pressure $P$ and specific internal energy $\epsilon$ are everywhere continuous and satisfy 
\begin{subequations}
\begin{eqnarray}
    P(\rho)&=& K_i\rho^{\Gamma_i}, \\
  \epsilon(\rho)&=& a_i+\frac{K_i}{\Gamma_i-1}\rho^{\Gamma_i-1},
\end{eqnarray} 
\end{subequations}
inside the range of rest-mass density $\rho_{i-1} \leq \rho < \rho_i$ ($i=1,2,3,4$).

\begin{figure}[t]
    \centering
    \includegraphics[width=\columnwidth]{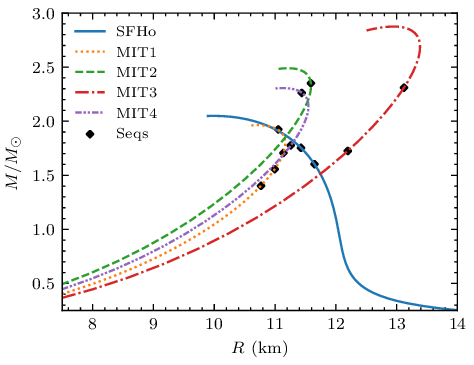}
    \caption{Gravitational mass is plotted against radius for nonrotating NSs modeled by the SFHo EOS and QSs modeled by four MIT bag models. 
    Points labeled by ``Seqs" correspond to the nonrotating configurations in the sequences of constant baryon mass or constant central density we used in the study of rotating stars. 
    }
    \label{fig:6MR}
\end{figure}

The parameters of our EOS models are summarized in Tables~\ref{table:eos1} and \ref{table:eos2} and the corresponding mass-radius relations for nonrotating QSs and NSs are shown in Fig.~\ref{fig:6MR}. 
The mass-radius relations of MIT bag models differ qualitatively from that of the SFHo EOS due to the well-known fact that QSs are self-bound objects. 

\subsubsection{For the atmosphere of QSs}

The surface of a bare QS modeled by the MIT bag model is identified by the vanishing pressure just like an ordinary NS, but it has a finite density $\rho_S$, which is of the same order as the central density, and it requires novel treatments in dynamical modelings. 
In traditional hydrodynamic simulations for NSs, a low-density atmosphere is introduced to fill up all vacuum space outside the stars, such that a fluid element is reset to become part of the atmosphere when its density evolves to become smaller than a prescribed value of the atmospheric density or even negative.   
This approach works well in highly dynamical situations such as inspiraling binary stars when the stars move across the computational grid in a short time scale, and the effects of the low-density atmosphere become relatively unimportant. 
However, in studying the oscillations of stable stars whose vacuum-fluid interface may move slowly, excessive oscillations may cause the star to extract (lose) mass from (to) the atmosphere and violate the conservation of mass and momentum. As the effects accumulate and amplify, they ultimately destabilize the evolution~\cite{Radice:2013xpa}. 
This situation only gets worse for QSs when the vacuum-fluid density discontinuities are many (ten) orders of magnitude larger than those in traditional NS cases. 
Immediately after starting the simulation, a large violation of mass conservation would be observed, and it soon rises up to the order of total mass, completely destroying the simulation.  

Furthermore, during a numerical simulation it can happen that fluid elements on the surface of a QS may evolve to a density smaller than the surface density $\rho_S$, which is defined by the vanishing pressure of the MIT bag model. 
As their densities ($\sim \rho_S$) are typically many orders of magnitude larger than the density of the atmosphere, they cannot simply be treated as part of the atmosphere. 
This then poses a question of how to evolve such fluid elements and with what type of EOS so that the dynamics near the surface of a QS can be modeled correctly. 

Instead of arbitrarily modifying some atmospheric elements during evolution, we want to maintain the balance between inertial and gravitational forces on all fluid elements and enforce the conservation law around the vacuum-fluid interface. 
In other words, a scheme allowing for a free evolution of the atmosphere is required.
To achieve this purpose, we introduce here a dustlike EOS to model fluid elements near the surface of a QS, whose rest-mass density is not necessarily small but whose pressure is always close to zero. 
Even though the truncation errors from finite differencing may cause a large density dislocation, its disturbance to the pressure profile would be minimal. 
The gravitational pull of the star tends to bring a dislocated fluid element back to its equilibrium position. 

In practice, after importing the initial data of a bare QS into the computational domain, an atmosphere with a floor rest-mass density $\rho_f$ is set outside the star. 
In our simulations, the value of $\rho_f$ is chosen to be $10^{-18}$, about $10^{-15}$ times the central rest-mass density of the star.    
During the evolution, we set a pressure cutoff or equivalently a density cutoff slightly larger than the surface density following our parametrization of the MIT bag model EOS: 
\begin{subequations}
\begin{eqnarray}
   \rho_{\rm{cutoff}}&=& \rho_S(1+\xi),\\
   P_{\rm cutoff}&\approx& B(1+c_{ss})\xi\approx c_{ss}\rho_{\rm cutoff} \xi,
\end{eqnarray}  
\end{subequations}
where $\xi$ is small. 
The specific enthalpy and energy are given by $h=(1+\xi)^{c_{ss}}\approx 1+ c_{ss} \xi$ and $\epsilon\approx c_{ss}\xi^2/2$, respectively.
It should be noted that the effective adiabatic index diverges near the stellar surface 
\begin{equation}
    \Gamma=\frac{d\ln P }{d\ln\rho} = \frac{(1+c_{ss})\kappa^{1+c_{ss}}}{\kappa^{1+c_{ss}}-1}\approx \frac{1}{\xi},
\end{equation}
meaning the surface is theoretically infinitely stiff. 
If the rest-mass density is 
above the cutoff density $\rho_{\rm cutoff}$, then we will use Eq.~(\ref{mit:eos}). Otherwise, we switch to the following EOS for the fluid element:
\begin{subequations}
\begin{eqnarray}
     P(\rho)&=& c_{ss} \xi \rho ,\\
    e(\rho)&=&  \left(1+\frac{c_{ss}}{2}\xi^2\right)\rho .
\end{eqnarray} 
\end{subequations}
When $\rho < \rho_{\rm cutoff}$, the transition of the EOS from the MIT bag model to dust does not mean a physical phase transition. 
The dust atmosphere can be considered as being motivated by the physical picture that when self-bound quark matter droplets, each with density $\approx \rho_{\rm cutoff}$, are ejected from the stellar surface in dynamical evolution, they form a thin layer of atmosphere. 
In a finite-volume cell, the number of droplets is not large enough to form a fluid, but instead well described by a system of pressureless dust.
Both the specific energy and enthalpy are enforced to be continuous across the surface discontinuity.
The first law of thermodynamics, which requires $de/d\rho=h$, is violated to the order of $\mathcal{O}(\xi)$.  
By choosing a small $\xi$, we expect our model to capture the dynamics near the surface of a bare QS, and eventually the error is dominated only by the finite-differencing error.      
We used $\xi=10^{-12}$ in our simulations. The cutoff pressure is then also about $10^{-12}$ of the central pressure, and the surface specific internal energy $\epsilon\propto\xi^2$ is below the roundoff precision of the double precision floating point numbers. 
As we shall see in the following, the introduction of this dustlike EOS near the surface of a QS enables us to determine the radial oscillation modes of QSs accurately.    

\subsection{Other numerical issues}

In addition to the Riemann solver (see Sec.~\ref{sec:ppsolver}), which determines the numerical flux at the cell interfaces, one also needs a reconstruction method to interpolate the fluid variables. 
In our simulations, we implemented the classic piecewise parabolic method (PPM) reconstruction method~\cite{COLELLA_1984}. 
It should be pointed out that the original PPM scheme applies a steepening procedure for density discontinuity only if the following condition is satisfied (see Eq.~(3.2) in \cite{COLELLA_1984}):
\begin{equation}
    \Gamma K_0 \frac{|\rho_{j+1}-\rho_{j-1}|}{\min{(\rho_{j+1},\rho_{j-1})}} \ge \frac{|P_{j+1}-P_{j-1}|}{\min{(P_{j+1},P_{j-1})}} , 
\end{equation}
where $K_0$ is a constant parameter. This condition determines whether the $j$th zone can be treated as being inside a discontinuity. However, this criterion does not work properly for QSs due to the divergence of the effective adiabatic indices $\Gamma\approx 1/\xi$ of QSs near the surface.
As a result, any pair of constants $\Gamma$ and $K_0$ in the PPM scheme would not properly detect discontinuities near the surface of a QS. 
In our simulations, we simply turned off this condition and always allowed the steepening procedure for QS models. 
This adjustment can sharpen surface density profiles and prolong our simulations.
We have also tested the fifth-order monotonicity preservation scheme (MP5)~\cite{SURESH199783,MIGNONE20105896} and found no advantage regarding mass conservation, as we shall discuss in Sec.~\ref{sec:TOVQS}. 

In this study, we do not attempt to extract the gravitational wave signals emitted from oscillating stars using the Newman-Penrose formalism, which is routinely employed in binary neutron star simulations. 
The outer boundary of the computational domain in our simulations can then be set closer to the stellar surface. 
Nevertheless, we found that the quality of the hydrodynamic modeling of a rapidly rotating QS, such as mass conservation, can be affected strongly if the outer boundary is too close to the stellar surface. 
In our simulations, we employ three refinement levels with a $2:1$ refinement ratio for successive levels provided by the mesh refinement driver \texttt{CARPET} in order to maintain enough grid resolution inside the star, while the outer boundary can be put far away from the stellar surface to reduce its effects.    
The first refinement boundary is at a radius of $1.2r_{\rm eq}$, the second at $2.4r_{\rm eq}$, and the outer boundary at $4.8r_{\rm eq}$, where $r_{\rm eq}$ is the equatorial coordinate radius. 
In Sec.~\ref{sec:TOVQS}, numerical results of three spatial resolutions $\Delta x=0.12\ (\approx 177\ {\rm m})$, $0.16\ (\approx 236\ {\rm m})$, and $0.24\ (\approx 354\ {\rm m})$, 
in the case of a nonrotating QS are compared, where $\Delta x$ is the grid size for the finest level. 
Since we could already extract the frequencies of radial oscillation modes of QSs up to the fifth overtone
using $\Delta x =0.24$, we produced our results with the default resolution $\Delta x=0.16$.
For a typical slowly rotating star, its radius will cover about $50$ computational grids. 
Fast rotation can cause a large deformation of the star, and the ratio between the polar and equatorial radii can be below $0.5$ for some extreme models. 
For these cases, there are about $30$ and $60$ cells along the polar and equatorial radii, respectively.  

To save computational resources, reflection symmetry about the equatorial plane is assumed and interesting modes ($l=|m|=2$)
will not be affected by this choice. 
Octant symmetry is applied when bar modes are not concerned. 
In short summary, our main numerical results in this work are obtained by using the PP Riemann solver with the PPM reconstruction method. The time update is performed using the standard RK4 integrator.

\subsection{Initial data and perturbations}

We use the numerical code \texttt{rotstar} in the \texttt{LORENE}~\cite{Loreneweb} library to construct uniformly rotating NS and QS models for our study. 
The code uses a multidomain spectral method \cite{Bonazzola:1993,Bonazzola:1998} and has been used for calculating rapidly rotating QSs \cite{Gourgoulhon_1999,Gondek2000,Gondek2003,Lo_2011}.  
Sequences of constant baryon mass and constant central energy density are produced, whose corresponding 
nonrotating configurations are labeled by ``Seqs" data points in Fig.~\ref{fig:6MR}.

\begin{figure}[t]
    \centering
    \includegraphics[width=\columnwidth]{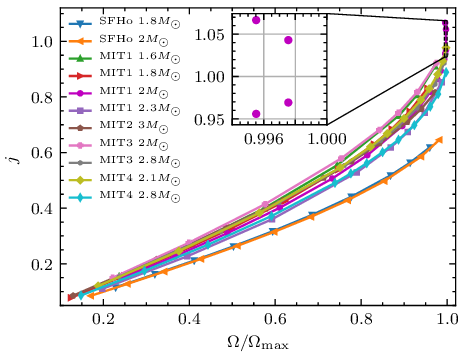}
    \caption{
The spin parameter $j$ is plotted against the angular velocity $\Omega$ normalized by the corresponding maximum rotation limit $\Omega_{\rm max}$ for constant baryon mass sequences.
In the figure legend, each sequence is labeled by the EOS model and the (fixed) value of the baryon mass of the sequence. 
For NSs, the maximal rotational frequency is its Keplerian limit, $\Omega_K = \Omega_{\rm max}$; but for QSs, $\Omega_{\rm max}$ is larger than $\Omega_K$ by about $2\%$~\cite{Gondek2000} typically.  
The inset enlarges the two pairs of degenerate models of the MIT1 $2M_\odot$ sequence in the sense that each pair has the same rotational frequency. 
The data set contains 115 models, including 21 SFHo models and 41 MIT1, 10 MIT2, 18 MIT3 and 25 MIT4 QS models.    }
    \label{fig:jfk}
\end{figure}

An important dimensionless parameter to characterize a rotating compact star is its spin parameter $j=J/M^2$, where $J$ and $M$ are the angular momentum and gravitational mass of the star, respectively. 
In Fig.~\ref{fig:jfk}, the spin parameters of our constant baryon mass sequences are plotted against the angular velocity $\Omega$, normalized by the corresponding maximal rotation limit $\Omega_{\rm max}$. 
There is a gap separating the band of QS models from the two NS sequences in the figure. 
For the same baryon mass, the spin parameters of QSs are significantly larger than the NS counterparts, especially when $\Omega / \Omega_{\rm max}$ approaches unity.   
It should be pointed out that, in contrast to the situation for rotating NSs, the maximum angular velocity $\Omega_{\rm max}$ of a sequence of QSs with a given baryon mass is generally higher than the Keplerian limit $\Omega_K$ by about $2\%$, a characteristic feature of self-bound objects that a QS can further gain angular momentum by slightly slowing down its rotation but increasing its oblateness (i.e., the moment of inertia) before reaching the Keplerian limit~\cite{Gondek2000}. 
While the two NS sequences with a $10\%$ difference in baryon mass match each other very well, the QS sequences for a given EOS model are seen to depend more sensitively on the mass. 
In particular, the spin parameter for QSs increases as the baryon mass decreases.  
As pointed out in~\cite{Lo_2011}, the spin parameter of QSs can even be larger than the Kerr bound $j=1$ for rotating black holes. 
We study two such models in the MIT1 $2M_\odot$ sequence, which are also degenerate models in terms of the rotational frequency, as shown in the inset of Fig.~\ref{fig:jfk}.
Clearly, the maximal rotational frequency is a turning point. 
On the other hand, there is an upper bound of $j \sim 0.7$ for uniformly rotating
NSs, the value of which is relatively insensitive to EOS models \cite{Lo_2011,Cipolletta:2015,Koliogiannis:2020}. 

\begin{figure}[t]
    \centering
    \includegraphics[width=\columnwidth]{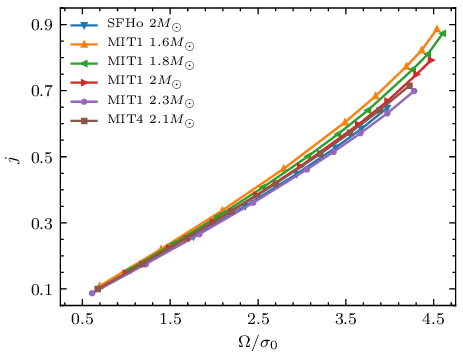}
    \caption{
    The spin parameter $j$ is plotted against the angular velocity $\Omega$ normalized by the $f$-mode frequency $\sigma_0$ of the corresponding nonrotating stars for constant central energy density sequences.  In the figure legend, each sequence is labeled by the EOS model and the baryon mass of the nonrotating star in the sequence. The data set contains 52 models, including 6 SFHo models and 37 MIT1 and 9 MIT4 QS models.
    }
    \label{fig:jf0}
\end{figure}

Similarly, the sequences of constant central energy density are plotted in Fig.~\ref{fig:jf0}. 
In the figure, we plot $j$ against the ratio $\Omega / \sigma_0$, where $\sigma_0$ is the $f$-mode frequency of the corresponding nonrotating star for each sequence. 
In contrast to Fig.~\ref{fig:jfk}, there is now no qualitative difference between the NS and QS, except that the latter can reach higher values of $j$ and $\Omega / \sigma_0$. 
The reason to normalize $\Omega$ by $\sigma_0$ is due to the fact that the $f$-mode frequencies of NSs for these sequences establish a universal relation with $\Omega / \sigma_0$ \cite{Kokkotas2020}. 
On the other hand, for the sequences of constant baryon mass, the $f$-mode frequencies of NSs observed in the rotating frame are connected to $\Omega / \Omega_K$ by another universal relation.
We shall study whether the $f$-modes of rotating QSs for these two types of sequences still satisfy the universal relations found for NSs.


Every data point in Figs.~\ref{fig:jfk} and \ref{fig:jf0} represents a rotating star model we perturbed and evolved dynamically to $t=2000\ (\approx 9.85\ {\rm ms})$, and their oscillation modes were then extracted for further analysis. 
To excite the quadrupolar nonaxisymmetric ($l=|m|=2$) oscillation modes of rotating stars, we add initial velocity perturbations following the suggestion of~\cite{Dimmelmeier2006}. 
After importing the initial data for an equilibrium rotating star to the evolution code, we perturb the star by adding the velocity perturbations
\begin{subequations}
\begin{eqnarray}
    v^\theta&=& v^r \sin 2\theta (\cos 2\phi+\sin2\phi), \\
       v^\phi &=&  -2v^r\sin\theta(\sin2\phi-\cos2\phi), 
\end{eqnarray}
\end{subequations}
where the radial component $v^r$ controls the perturbation strength, which we set to be $v_0\sin[\pi r/2r_s(\theta)]$ for some small values of $v_0$, and $r_s(\theta)$ is the estimated coordinate radius along the $\theta$ direction. 
A typical choice for $v_0$ used in our simulations is 0.005, which contributes negligibly to the initial Hamiltonian constraint violation comparing to the numerical error from importing and interpolating the \texttt{LORENE} initial data to the \texttt{CACTUS} Cartesian grids.
Although these perturbation functions are not the exact eigenmodes of rapidly rotating stars, they can effectively excite the fundamental $f$-modes and also the first pressure $p$-modes.

\section{Numerical results} 
\label{sec:TestAndResult}

\subsection{Nonrotating QSs}
\label{sec:TOVQS}

Before studying the oscillation modes of rotating QSs, let us first present various tests for nonrotating models to demonstrate that our numerical method is capable to provide a stable and accurate evolution for a bare QS.   
In particular, we focus on a nonrotating QS with a gravitational mass $1.71 M_\odot$ and a radius $11.1$ km described by the MIT1 EOS. 
This star corresponds to the nonrotating configuration of a constant baryon mass ($2 M_\odot$) sequence in our study of rotating QSs.  

\begin{figure}[t]
    \centering
    \includegraphics[width=\columnwidth]{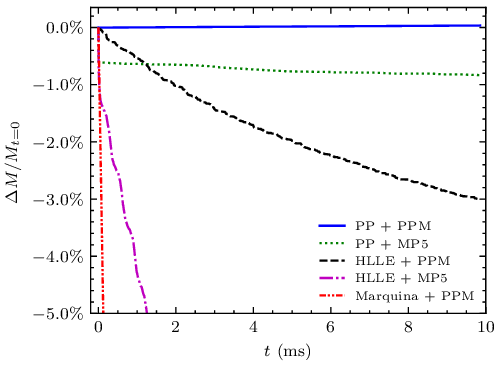}
    \caption{Evolutions of the percentage changes of the total baryon mass of a nonrotating QS for five different combinations of the Riemann solvers and reconstruction methods using a medium grid resolution $\Delta x = 0.16\ (\approx 236 \ {\rm m})$ at the finest refinement level.     
    The default scheme PP+PPM employed in this study can preserve the mass conservation to high accuracy, about 0.034\% at $t\approx 9.85$ ms. 
    }
    \label{fig:whypp}
\end{figure}

As discussed before, we choose the PP Riemann solver with PPM reconstruction method 
(hereafter PP+PPM) as our default hydrodynamic scheme. 
Here we first compare the performance of the PP+PPM scheme with other standard Riemann solvers, the Harten-Lax-van Leer-Einfeldt (HLLE)~\cite{HartenLaxLeer1983,Einfeldt1988} and Marquina \cite{DONAT199642,Aloy_1999} schemes, and MP5~\cite{SURESH199783,MIGNONE20105896} reconstruction method. 
In Fig.~\ref{fig:whypp}, we plot the percentage changes of total baryon mass against time for the simulations 
using different combinations of the Riemann solvers and reconstruction methods. 
The simulations were performed with the same grid resolution $\Delta x=0.16$ at the finest refinement level. 
It is seen that the Marquina+PPM scheme loses a large percentage of mass immediately at the beginning of the evolution. 
Similarly, the HLLE+MP5 scheme also has a large decrease in mass initially and the mass loss increases to 5\% by about 1.3 ms. 
Replacing the MP5 method with a lower order (third-order) PPM method can improve the mass conservation as can be seen by comparing the HLLE+MP5 and HLLE+PPM schemes in the figure. 
The HLLE+PPM scheme gradually loses 3\% of total mass by about 10 ms. 
In fact, we noticed that a higher-order reconstruction method actually causes more spurious oscillations near the sharp surface discontinuity. 
By comparison, it is seen clearly that our implemented PP solver, whether using the MP5 or PPM reconstruction method, performs much better than the other solvers and can conserve the total mass to within 1\% up to 10 ms. 
While the PP+MP5 run still suffers an initial drop in mass, the PP+PPM scheme is nearly a flat line. 
The numerical results presented in this paper are obtained by the PP+PPM scheme hereafter.     

\begin{figure}[t]
    \centering
    \includegraphics[width=\columnwidth]{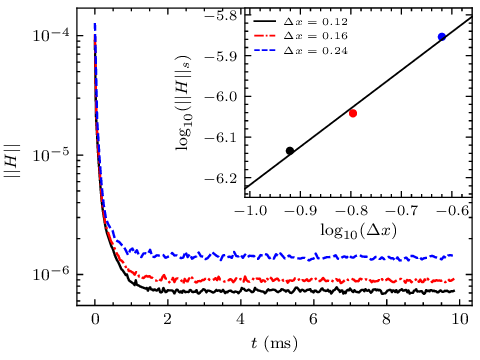}
    \caption{$L^2$-norm of the Hamiltonian constraint violation $||H||$ in the evolutions of a nonrotating QS for three different grid resolutions. 
    The inset plots the stable plateau values $||H||_s$ of the constraint violation and demonstrates linear-order convergence for $||H||_s$. 
    }
    \label{fig:tovham}
\end{figure}

An important quantity to monitor the quality of a numerical-relativity simulation is the Hamiltonian constraint. 
A small constraint violation is required for any trustworthy simulation. 
In Fig.~\ref{fig:tovham}, we plot the $L^2$ norm of the Hamiltonian constraint against time for the evolution of the nonrotating QS using three different resolutions $\Delta x = 0.24$ (low), $0.16$ (medium), and $0.12$ (high). 
Thanks to the constraint damping and propagation properties of the CCZ4 formulation, the Hamiltonian constraint violation quickly drops to a steady plateau of $\mathcal{O}(10^{-6})$, 
2 orders of magnitude smaller than the initial Hamiltonian constraint violation, even in the low-resolution run. 
The figure shows that the violation decreases with increasing resolution. The inset of Fig.~\ref{fig:tovham} plots the stable plateau values $||H||_s$ of the constraint violation against $\Delta x$, and demonstrates a linear-order convergence for $||H||_s$. 


After checking the stability and accuracy of the evolution, we now turn to the oscillations of the nonrotating QS model. 
While the star is a static equilibrium configuration initially, finite-differencing errors can trigger the radial oscillation modes during the evolution. 
The frequencies of the oscillation modes can then be obtained by performing Fourier transforms (FT) of physical quantities such as the density and velocity.
For nonrotating stars, the oscillation mode frequencies can alternatively be computed using a perturbative eigenmode analysis. 
For radial oscillation modes, we followed~\cite{Kokkotas2001}; for nonradial quadrupolar modes, we followed~\cite{Lindblom1983,Detweiler1985ApJ,Lau2010}.
Comparing the mode frequencies obtained from the simulation with the known eigenmode frequencies is an important test of the hydrodynamic simulation. 

For a general rotating star, we use the \texttt{LINEOUT} module in the open source visualization and data analysis software \texttt{VisIt}~\cite{visitweb} to extract the physical quantities inside the star at data points along the line at polar angle $\theta=\pi/4$ on the $x$-$z$ plane, where $z$ is the rotation axis. 
The Fourier transforms of the rest-mass density $\rho$ at different data points are added up and the absolute value of their sum defines our final Fourier spectrum of $\rho$. 
Similarly, we also consider the Fourier spectra of the velocity components defined
by $v^r=(v^x+v^z)/\sqrt{2}$, $v^\theta = (v^x-v^z)/\sqrt{2}$, and $v^\phi = v^y$, where $(v^x , v^y , v^z$) are the velocity components obtained in our Cartesian grid simulations. 
In practice, we found that the Fourier spectrum of a physical quantity obtained from the superposition of multidata points could improve the quality of the spectrum and was helpful in the mode identification. 

\begin{figure}[t]
    \centering
    \includegraphics[width=\columnwidth]{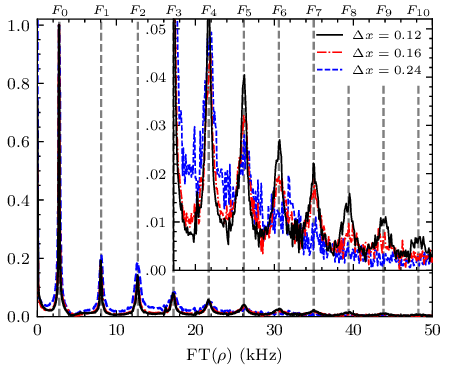}
    \caption{
    Fourier spectra of the rest-mass density for the evolutions of a nonrotating QS using three different grid resolutions. 
    The vertical dashed lines indicate the frequencies of the radial oscillation modes, from the fundamental $F_0$ mode to the tenth overtone $F_{10}$, 
    determined by the perturbative normal-mode analysis. 
    The inset enlarges the region between $F_3$ and $F_{10}$. 
   }
    \label{fig:tovspectrum}
\end{figure}

Let us first study the radial oscillation modes of the nonrotating QS model discussed above in Fig.~\ref{fig:tovham} as a test for our simulations. 
In Fig.~\ref{fig:tovspectrum}, we show the Fourier spectra of density ${\rm FT}(\rho)$ obtained from the evolutions using three different grid resolutions. 
The vertical dashed lines in the figure stand for the frequencies of the radial oscillation modes, ranging from the fundamental mode $F_0$ to the tenth overtone $F_{10}$, determined by the perturbative method as in~\cite{Kokkotas2001}.  
It is seen that our simulation results can produce Fourier peaks matching the dashed lines very well.
The amplitudes of the spectra are dominated by the fundamental mode $F_0$ as expected. 
Higher overtones with much smaller amplitudes are still identifiable in the spectra. 
The Cartesian grid cannot match the stellar surface exactly, and hence many overtones can be excited due to numerical perturbations near the surface. 
Being able to identify the high-frequency overtones would be a good criterion for a proper simulation of a stable QS. 
 
In order to show clearly the Fourier peaks of the high overtones, we enlarge the frequency range from the $F_3$ to $F_{10}$ overtones in the inset of Fig.~\ref{fig:tovspectrum}. 
The high-resolution result (black curve) aligns very well with all overtones up to the ninth overtone with frequency $F_9=43808$ Hz. 
Near the tenth overtone with frequency $F_{10}=48215$ Hz, a small bump still exists at the correct position. 
It is seen that the peaks of the medium-resolution result (red curve) can also match up to the ninth overtone.
However, the low-resolution result (blue curve) can only recover up to the fifth overtone with frequency $F_5=26138$ Hz as the run suffers from more numerical dissipation.

\begin{figure}[t]
    \centering
    \includegraphics[width=\columnwidth]{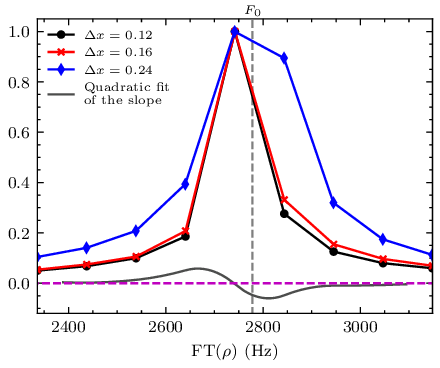}
    \caption{Spectra around the first peak in Fig.~\ref{fig:tovspectrum}. 
    The smooth black curve (without data points) represents the slope of a quadratic curve fit to the high-resolution result
    ($\Delta x = 0.12$). 
    The fundamental mode frequency $2745$ Hz obtained by the position at which the slope passes through zero agrees to the known normal-mode value (indicated by the vertical dashed line at $F_0 =2778$ Hz) to about 1.2\%. 
    }        
    \label{fig:F0interp}
\end{figure}

Let us end this subsection by discussing how we determined the mode frequencies from the Fourier spectra quantitatively. 
Obtaining accurate mode frequencies from a Fourier spectrum is important to our study of the universal relations of rotating QSs. 
In Fig.~\ref{fig:tovspectrum}, the QS is evolved to $t\approx 9.85$ ms for each resolution run and the resolution in the frequency of the Fourier spectrum is inversely proportional to this evolution time, meaning that the frequency resolution is $101.5$ Hz. 
Figure~\ref{fig:F0interp} is the same plot of ${\rm FT}(\rho)$ as Fig.~\ref{fig:tovspectrum}, but focuses around the fundamental $F_0$ mode. 
It is seen that the width of the peak decreases as the resolution increases, and the medium (red line) and high (black line) resolution results agree very well. 
To extract the mode frequency from the high-resolution result, we fit a quadratic curve around the peak, and the mode frequency is approximated by the position at which the slope of the curve passes through zero as illustrated by the smooth solid line in Fig.~\ref{fig:F0interp}.  
We obtain the fundamental mode frequency $2745$ Hz from the high-resolution run using this method, which differs from the known normal-mode value $F_0=2778$ Hz by about 1.2\%.  

In general, the radial oscillation modes are sensitive to the stellar profile, and the capability of our simulations to recover the correct mode frequencies to high overtones accurately suggests that we have modeled the sharp surface of the QS properly. 

\subsection{Stability of rotating QSs}

\begin{figure}[t]
    \centering
    \includegraphics[width=\columnwidth]{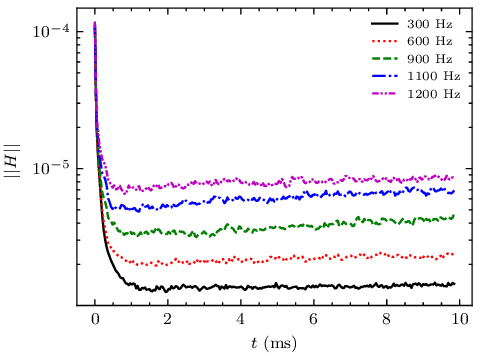}
    \caption{$L^2$-norms of the Hamiltonian constraint violations $||H||$ for a sequence of $2M_\odot$ baryon mass MIT1 QSs are plotted against time. The rotational frequencies of the chosen models span from
    300 to 1200 Hz, where the maximal rotation limit of the sequence is near $1228$ Hz. 
    The results are obtained using the same resolution $\Delta x =0.16$. 
     }
    \label{fig:rotham}
\end{figure}

To demonstrate that we can perform stable and accurate simulations of rapidly rotating QSs, we first show in Fig.~\ref{fig:rotham} the $L^2$-norm of the Hamiltonian constraint violations for a sequence of $2M_\odot$ baryon mass MIT1 QSs with rotational frequencies ranging from 300 to 1200 Hz, where the maximal rotation limit is near 1228 Hz. 
The runs were performed with the same grid resolution $\Delta x = 0.16$, which is the default resolution we used for obtaining the oscillation modes of rotating stars. 
Similar to what we have seen for nonrotating QSs, the constraint violations quickly drop to stable plateau levels at the beginning of the simulations and remain flat until $t \approx 9.85$ ms for all models including the most rapidly rotating one. 
Although the stable plateau values of the constraint violation get larger for faster rotation, it still maintains a relatively small value below $10^{-5}$, an order of magnitude smaller than the initial Hamiltonian constraint violation, and does not grow noticeably even for the $1200$ Hz model, which is close to the maximal rotation limit of the sequence. 

\begin{figure*}[ht]
    \centering
    \includegraphics[width=\textwidth]{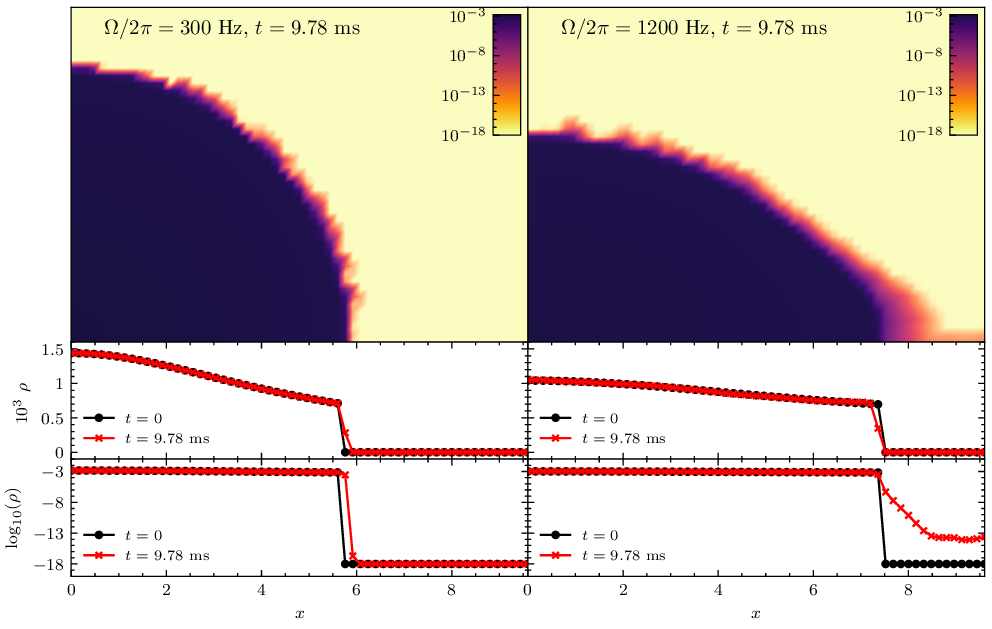}
    \caption{
    Snapshots of the rest-mass density in the first quadrant of the $x$-$z$ plane for two MIT1 QS models with rotational frequencies 300 Hz (top left) and 1200 Hz (top right).
    The density profiles of the two models along the $x$ axis are plotted in linear (middle panels) and logarithmic (bottom panels) scales.      
   }
    \label{fig:rotstar_profiles}
\end{figure*}

One important challenge for us is to demonstrate our ability to simulate the sharp surface of rapidly rotating QSs for a long duration.  
Figure~\ref{fig:rotstar_profiles} compares the snapshots of density profiles at $t\approx 9.78$ ms for the 300 Hz and 1200 Hz rotating QSs considered in Fig.~\ref{fig:rotham}. 
The top panels in the figure show the rest-mass densities of the stars in the first quadrant of the $x$-$z$ plane, where the $z$ axis is the rotation axis.   
The large color contrast from the large density gradient and the imperfect matching of the Cartesian grids to the star surfaces result in visible serrate edges at the surfaces.
While the slowly rotating 300 Hz model (left panel) still maintains a spherical shape very well, the 1200 Hz model (right panel) is flattened at the pole and develops an oblate shape due to rapid rotation. 
It can be seen that some tiny amount of matter is ejected from the surface under the influence of centrifugal force near the equatorial plane. 
Nevertheless, the baryon mass of this rapidly rotating model remains very well conserved to within 0.1\% error by the end of the simulation at $t\approx 9.85$ ms, which is equivalent to about 12 rotation periods. 
This mass-shedding effect would unavoidably occur for rapidly rotating models close to the Keplerian limit. 
The effect would dampen the stellar pulsations and the amplitudes of the oscillation modes would gradually decrease for rapidly rotating stars~\cite{Stergioulas2004}. 
It also affects the sharpness of the surface discontinuity near the equator. 
The dislocation of mass elements from their balance positions caused by the truncation brings constant disturbances to the stars and excites many oscillation modes. 

To clearly demonstrate the sharpness of the stellar surface, the middle and bottom panels in Fig.~\ref{fig:rotstar_profiles} show the density profiles of the two models along the $x$ axis at $t=0$ and $t\approx 9.78$ ms on the equatorial plane in linear (middle panels) and logarithmic (bottom panels) scales, respectively. 
The MIT1 EOS has a surface density $\rho_S \approx6.93\times10^{-4}$ in the code units, which drops to the floor density $\rho_f=10^{-18}$ in the slowly rotating model over one cell, but only drops 3 orders of magnitude for the rapidly rotating model due to the mass-shedding effect. 
Along other directions, the surface density would drop to the floor density over one or two cells.

\begin{figure}[t]
    \centering
    \includegraphics[width=\columnwidth]{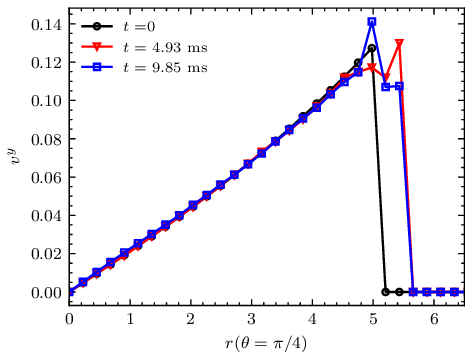}
    \caption{Plot of the velocity profile $v^y$ along the direction $\theta=\pi/4$ for the 1200 Hz rotating model studied in Fig.~\ref{fig:rotstar_profiles} at $t=0$, 4.93 ms, and 9.85 ms on the $x$-$z$ plane.
    }
    \label{fig:vyprofiles}
\end{figure}

To check the stability of the rotational velocity profile, we plotted $v^y$ along the $\theta=\pi/4$ direction on the $x$-$z$ plane ($\phi=0$) in Fig.~\ref{fig:vyprofiles} for the 1200 Hz rapidly rotating model. 
The profiles at $t=0$, 4.93 ms, and 9.85 ms overall agree very well, though small oscillations of the star surface across four grid cells can be seen. 
Figures~\ref{fig:rotstar_profiles} and \ref{fig:vyprofiles} clearly demonstrate the stability of the density and velocity profiles of rapidly rotating QSs in our simulations. 
In particular, the sharp density jump at the stellar surface can also be maintained very well.

\subsection{Oscillation modes of rotating QSs}

\begin{figure*}[ht]
    \centering
    \includegraphics[width=1.0\textwidth]{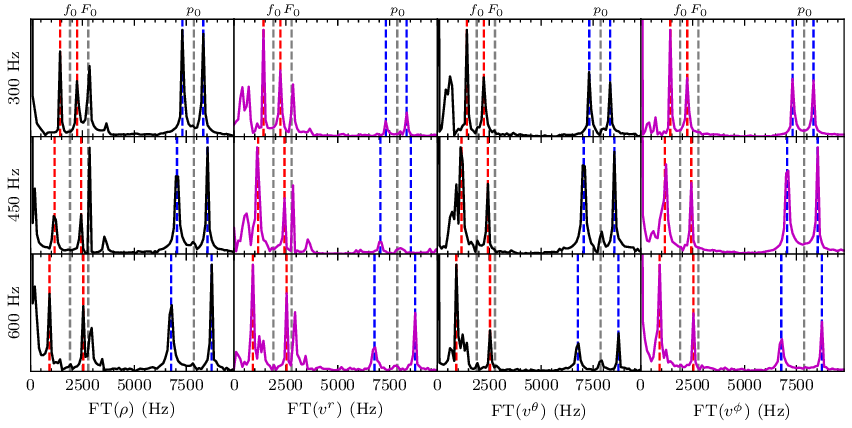}
    \caption{Fourier spectra of fluid variables $\rho$, $v^r$, $v^\theta$, and $v^\phi$ of the sequence of MIT1 QSs with constant baryon mass $2M_\odot$ rotating at $(300,450,600)$ Hz. 
    The gray dashed lines label the quadrupole fundamental mode $f_0=1897$ Hz, the first pressure mode $p_0=7868$ Hz, and the fundamental quasiradial mode $F_0=2778$ Hz of the corresponding nonrotating model. In each panel, two red (blue) dashed lines track the $m=\pm2$ $f$-modes ($p$-modes). 
    }
    \label{fig:combinespectrum_splitting}
\end{figure*}

In this section, we will focus on a sequence of MIT1 QS models with the same constant baryon mass $2M_\odot$, but different rotational frequencies. 
The sequence can be considered as a quasiequilibrium evolution of a rapidly rotating QS being slowed down to lower rotational frequencies if angular momentum is effectively transported away. 
By studying the Fourier spectra of the fluid variables of these stars, such as the rest-mass density 
$\rho$ and three-velocity components $v^r$, $v^\theta$, and $v^\phi$, we can extract their oscillation
mode frequencies.

\subsubsection{Fourier spectra and mode selectivity}

In perturbation theory as in~\cite{Detweiler1985ApJ}, when expanded in spherical harmonics $Y_{lm}$, each oscillation mode is associated to a pair of indices $(l, m)$. For a spherical nonrotating star, the different 
orders of $m$ are degenerate for a given $l$, and it is enough to consider the $m=0$ mode. 
For the $l=2$ quadrupolar modes that we focus on in this work, the degeneracy is broken by rotation and the bar modes ($m=\pm2$) split from the axisymmetric ($m=0$) mode, similar to the Zeeman effect in quantum mechanics.  
This phenomenon is clearly observed in Fig.~\ref{fig:combinespectrum_splitting} which shows the Fourier
spectra of density ${\rm FT}(\rho)$ and velocity components ${\rm FT}(v^r)$, ${\rm FT}(v^\theta)$, and ${\rm FT}(v^\phi)$ 
for QS models with rotational frequencies 300 Hz (first row), 450 Hz (middle row), and 600 Hz 
(bottom row) of the chosen sequence. The positions of the $f$-mode ($f_0 =1897$ Hz), the first pressure mode ($p_0=7868$ Hz), and the fundamental radial mode ($F_0 = 2778$ Hz) for the nonrotating configuration of the sequence are labeled by the 
gray dashed lines. 
In each panel, the $f_0$ and $p_0$ gray lines are each sandwiched by two sharp Fourier peaks labeled
by the red and blue dashed lines, respectively. The separation between the red (blue) lines decreases and converges to the $f_0$ ($p_0$) gray line as the rotation rate decreases towards zero. These peaks are 
the nonaxisymmetric $m=\pm2$ modes split from the $f_0$ and $p_0$ modes. The peak labeled by the 
left red line (and similarly for the left blue line) is the counterrotating $m=2$ mode, while the
right red line is the corotating $m=-2$ mode.       
Our initial velocity perturbation is chosen to excite the $m=\pm2$ modes strongly, but not the 
axisymmetric $m=0$ modes. However, as the rotation rate increases, small peaks corresponding to the $m=0$ modes near the positions of $f_0$ and $p_0$ start to appear. 

The fundamental quasiradial mode is also strongly excited in our simulations as can be seen by the 
peaks near the $F_0$ lines. It is not so surprising as radial oscillation modes can be easily excited 
due to finite-differencing errors as we have already seen for nonrotating QSs. 
The frequency of the quasiradial mode increases slightly with the rotation rate as can be seen from the spectra of $\rho$ and $v^r$. 
Nevertheless, it is still well approximated by its nonrotating counterpart $F_0$ even for the model rotating at 600 Hz, as the ratio between the polar and equatorial radii of this star is about 0.92 and the rotation effect is relatively small.     

Another interesting feature of the spectra shown in Fig.~\ref{fig:combinespectrum_splitting} is 
a selective effect of the appearance of different modes and their amplitudes in different spectra. For instance, the fundamental quasiradial mode establishes strong peaks in the spectra of $\rho$ and $v^r$, but not for $v^\theta$ and $v^\phi$, which may already be expected. 
Similarly, the peaks associated to the $m=0$ $p$-mode can be observed in the spectra of $v^\theta$ for the 450 Hz and 600 Hz models, while the corresponding peaks in the other spectra have much smaller 
amplitudes. 


\subsubsection{Onsets of secular instabilities}

\begin{figure*}[ht]
    \centering
    \includegraphics[width=\textwidth]{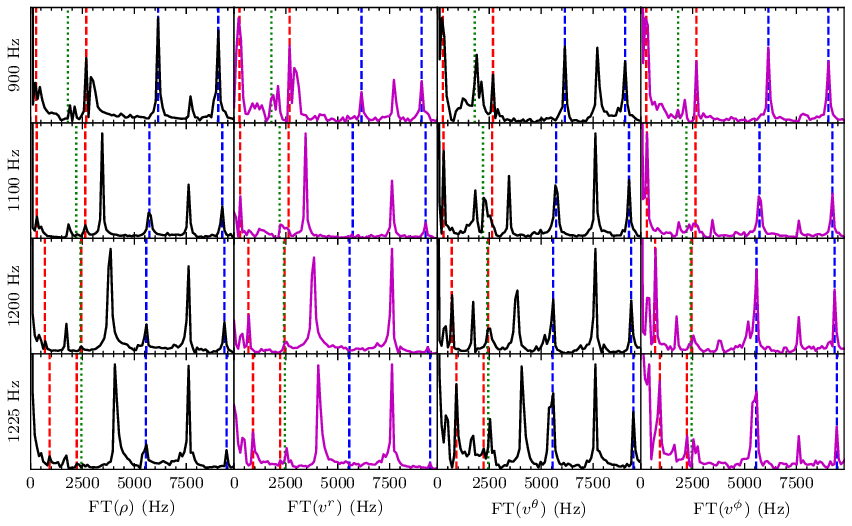}
    \caption{Fourier spectra of fluid variables $\rho$, $v^r$, $v^\theta$, and $v^\phi$ of the sequence of MIT1 QSs with constant baryon mass $2M_\odot$ rotating at $(900,1100,1200,1225)$ Hz. 
    Similar to Fig.~\ref{fig:combinespectrum_splitting}, the red and blue dashed lines track the 
    $m=\pm 2$ $f$- and $p$-modes. 
    The green dotted line in each panel tracks the position of twice the rotation frequency of the star. 
    The maximal rotation rate of this sequence is about 1228 Hz. See text for the identification of 
    the onsets of CFS and viscosity-driven instabilities from the spectra. 
    }
    \label{fig:combinespectrum_instability}
\end{figure*}

As the rotation rate increases, the peaks of the interesting $m=\pm2$ bar modes become less distinct and their amplitudes can even be smaller than the $m=0$ modes in some of the Fourier spectra.
Figure~\ref{fig:combinespectrum_instability} plots the Fourier spectra of the same sequence as in 
Fig.~\ref{fig:combinespectrum_splitting}, but for four models with higher rotational frequencies
up to 1225 Hz, which is very close to the maximum rotational frequency (1228 Hz) of this sequence.
In each panel, the red (blue) lines still track the $m=\pm2$ $f$-mode ($p$-mode), though the gray lines for $f_0$, $F_0$, and $p_0$ are not shown. The green line tracks the position of twice the rotation frequency of the star and its role will be explained below. 

It is clear that the Fourier spectra in Fig.~\ref{fig:combinespectrum_instability} shows some qualitative differences comparing to those for the slower rotating models considered in
Fig.~\ref{fig:combinespectrum_splitting}. First of all, starting from the 1100 Hz model, the fundamental quasiradial mode now has large amplitudes not only in the spectra of $\rho$ and $v^r$, but also those of $v^\theta$ as can be seen from the large peaks between the red and blue dashed lines in these spectra.  
As the rotation rate and oblateness of the star increase, the quasiradial modes couple $v^\theta$ and $v^r$, however, this coupling only becomes strong when the rotation rate is above 1000 Hz, which is 
close to the maximal rotation rate 1228 Hz of this sequence. 
In addition, the axisymmetric $m=0$ $f$- and $p$-modes are also excited to relatively large amplitudes
comparing to the case for slower rotating models. 
By tracking the mode positions and comparing the amplitudes in different spectra, the $m=\pm 2$ modes
can still be identified.
In contrast to Fig.~\ref{fig:combinespectrum_splitting}, the $m=0$ $p$-mode, which is identified to be the peak between the two blue lines in each panel, establishes larger amplitudes than the $m=\pm2$
counterparts (blue lines) in the $\rho$, $v^r$, and $v^\theta$ spectra when the rotation frequency is above 1000 Hz. However, the frequencies of the $m=0$ modes are not sensitive to the rotation rate.

Let us now focus on the $m=\pm 2$ $f$-mode (red dashed lines) and see how the onsets of secular instabilities for them are identified. 
As already been seen in Fig.~\ref{fig:combinespectrum_splitting}, the frequency of the counterrotating $m=2$ mode, which is tracked by the left red line in each panel, decreases as the rotation rate 
increases. However, further increasing the rotation rate from 900 Hz as illustrated in 
Fig.~\ref{fig:combinespectrum_instability} will push the mode to cross zero and become negative. 
Since the Fourier spectrum has even symmetry, the counterrotating mode appears to be ``reflected"
by the zero point and then shifts towards the right. The reflection occurs when the rotation frequency is at about 1000 Hz, which stands for the onset of the CFS instability (see Sec.~\ref{sec:intro})
for this sequence.  

For the $m=-2$ corotating mode, which is tracked by the right red line in each panel, its frequency increases initially along the sequence and then starts to decrease when the rotation rate increases above 900 Hz. We find that this sequence passes the viscosity-driven instability point 
(see Sec.~\ref{sec:intro}) when the rotation rate is about 1200 Hz. This instability sets in
when the frequency $\sigma_c$ of the corotating mode in the rotating frame goes through zero. 
Since $\sigma_c$ is related to the inertial-frame mode frequency $\sigma_i$ and the angular  
velocity $\Omega$ of the star by $\sigma_c = \sigma_i + m \Omega/2 \pi$, the instability sets in 
when $\sigma_i = 2 \Omega/(2\pi)$ (for $m=-2$). In Fig.~\ref{fig:combinespectrum_instability}, 
the quantity $2\Omega/(2\pi)$ is tracked by the green line in each panel, and hence the instability 
sets in when the right red line crosses the green line, as illustrated in the 1200 Hz model in the 
figure. 
As pointed out in Sec.~\ref{sec:intro}, the viscosity-driven instability of rotating QSs was 
studied before by perturbing the stellar configuration during the iteration steps in the construction of 
an axisymmetric equilibrium rotating star \cite{Gourgoulhon_1999,Gondek2000,Gondek2003}. 
Our study represents the first investigation based on the analysis of the oscillation modes.

\subsection{Universal relations of $f$-modes}
\subsubsection{Comparison to the universal relations for NSs}
\label{sec:cmpkk}

\begin{figure}[t]
    \centering
    \includegraphics[width=\columnwidth]{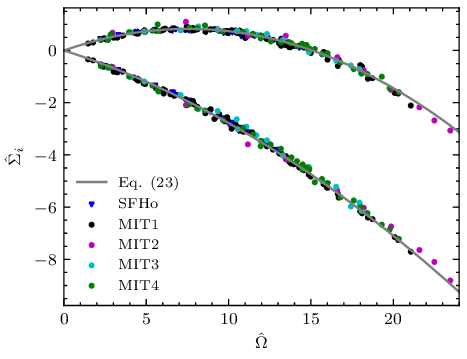}
    \caption{Plot of $\hat{\Sigma}_i$ [see Eq.~(\ref{eq:Sigma_i})] against the scaled angular velocity 
    $\hat\Omega$ for a total of 167 star models, including 27 SFHo NSs, and 78 MIT1, 10 MIT2, 18 MIT3,
    and 34 MIT4 QSs. The predictions from Eq.~(\ref{eq:kk3}) 
    (see also Eq.~(6) in \cite{Kokkotas2020}) for the counterrotating and corotating $f$-modes are given   
    by the lower and upper gray lines, respectively.}
    \label{fig:KK3rdUniversal}
\end{figure}

KK \cite{Kokkotas2020} recently proposed three universal relations for the $l=|m|=2$ $f$-modes
of rapidly rotating NSs. Here we shall study whether rapidly rotating QSs also satisfies these relations. 
We first compare our extracted mode frequencies from a total of 161 rotating NS and QS models with their
relation given by Eq.~(6) in \cite{Kokkotas2020},  
which relates the scaled mode frequency $\hat{\sigma}_i\equiv\bar{M}\sigma_i/\rm{kHz}$ in 
the inertial frame to the scaled angular velocity $\hat{\Omega}\equiv\bar{M}\Omega/\rm{kHz}$
and the effective compactness $\eta_{45}\equiv\sqrt{\bar{M}^3/I_{45}}$ by 
\begin{equation}
\label{eq:kk3}
    \hat{\sigma_i} = \left(c_1+c_2\hat{\Omega}+c_3\hat{\Omega}^2\right) + \left(d_1+d_3\hat{\Omega}^2\right)\eta_{45},
\end{equation} 
where $\bar{M}\equiv M/M_\odot$ and $I_{45}\equiv I/(10^{45}\ \rm g\cdot{cm}^2)$ are the star's scaled gravitational mass and moment of inertia.
The fitting coefficients $c_i$ and $d_i$ are given by 
$(c_1,c_2,c_3)=(-2.14,-0.201,-7.68\times10^{-3})$ and $(d_1,d_2,d_3)=(3.42,0,1.75\times10^{-3})$
for the counterrotating branch. For the corotating branch, 
$(c_1,c_2,c_3)=(-2.14,0.220,-14.6\times10^{-3})$ and $(d_1,d_2,d_3)=(3.42,0,6.86\times10^{-3})$. 
As each branch of data lies on a surface in the three dimensional 
$\eta_{45}$-$\hat{\sigma}$-$\hat{\Omega}$ parameter space, to have a clear visualization of the data, 
we define
\begin{equation}
\hat{\Sigma}_i \equiv \hat{\sigma}_i-c_1-(d_1+d_3\hat{\Omega}^2)\eta_{45} ,
\label{eq:Sigma_i}
\end{equation}
and plot it against $\hat{\Omega}$ in Fig.~\ref{fig:KK3rdUniversal}. 
In the figure, the lower (upper) branch of data consists of the counterrotating (corotating) modes. 
The predictions from Eq.~(\ref{eq:kk3}), which is Eq.~(6) in \cite{Kokkotas2020}, are labeled 
by the gray lines. It is noted that the nuclear matter SFHo EOS was not used in \cite{Kokkotas2020}, 
and hence our NS data can serve as an independent check for the universal relation. It is seen
that the $f$-modes of rapidly rotating QSs can also be described by this relation very well. 
The root-mean-square of the residuals is $0.111$ for the counterrotating branch and $0.0897$ for the 
corotating branch.


\begin{figure}[t]
    \centering
    \includegraphics[width=\columnwidth]{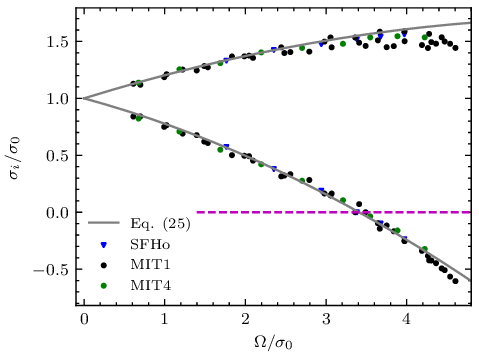}
    \caption{
    Plot of the scaled mode frequencies $\sigma_i/\sigma_0$ observed in the inertial frame 
    for sequences of constant central energy density. Data points contain the same models as in Fig.~\ref{fig:jf0}. The predictions from Eq.~(\ref{eq:kk1}) (see also Eq.~(4) in \cite{Kokkotas2020}) for the counterrotating and corotating $f$-modes are given by the lower and upper gray lines, respectively. The purple horizontal dashed line represents the zero-frequency line, on which the counterrotating mode becomes unstable to the CFS instability.    }
    \label{fig:KK1stUniversal}
\end{figure}

We next examine another universal relation for the mode frequency $\sigma_i$ observed in the inertial frame for sequences of constant central energy density, given by Eq.~(4) in \cite{Kokkotas2020},
\begin{equation}
\label{eq:kk1}
    \frac{\sigma_i}{\sigma_0} = 1+a_1 \left(\frac{\Omega}{\sigma_0}\right)+a_2\left(\frac{\Omega}{\sigma_0}\right)^2 ,
\end{equation}
where $(a_1,a_2)=(-0.193,-0.0294)$ for the counterrotating branch and $(a_1,a_2)=(0.220,-0.0170)$ for the corotating branch, and the angular velocity $\Omega$ is normalized by the $f$-mode frequency 
$\sigma_0$ of the corresponding nonrotating star. 
Our extracted mode frequencies also match closely to Eq.~(\ref{eq:kk1}) as shown in Fig.~\ref{fig:KK1stUniversal}.
The root-mean-square of the residuals is $0.0341$ for the counterrotating branch and $0.0794$ for the 
corotating branch. It is noted that the data points for the corotating branch (i.e., the upper branch in 
Fig.~\ref{fig:KK1stUniversal}) have larger deviations from Eq.~(\ref{eq:kk1}) for high rotation rates
close to the Keplerian limit, the region where Eq.~(\ref{eq:kk1}) does not fit well even for NS data 
as can be seen from Fig.~1 in \cite{Kokkotas2020}. 
The purple horizontal dashed line represents the zero-frequency line, on which the counterrotating
mode becomes unstable to the CFS instability.  
We find that QSs become unstable when the rotation rate $\Omega \approx 3.4\sigma_0$, which 
agrees with the finding for NSs \cite{Kokkotas2020}.   

\begin{figure}[t]
    \centering
    \includegraphics[width=\columnwidth]{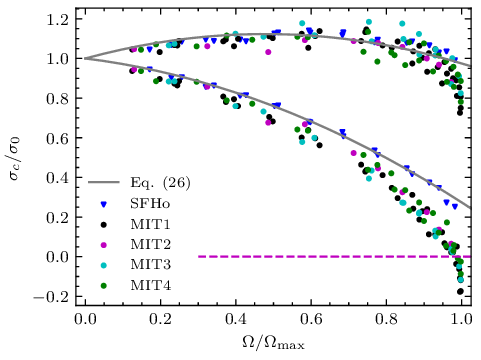}
    \caption{Plot of the scaled mode frequencies $\sigma_c /\sigma_0$ observed in the rotating frame
    for sequences of constant baryon mass. Data points contain the same models as in Fig.~\ref{fig:jfk}. The predictions from Eq.~(\ref{eq:kk2}) (see also Eq.~(5) in \cite{Kokkotas2020}) for the counterrotating and corotating $f$-modes are given by the upper and lower gray lines, respectively. The purple horizontal dashed line represents the 
zero-frequency line, on which the corotating mode becomes unstable to the viscosity-driven instability. 
    }
    \label{fig:KK2ndUniversal}
\end{figure}

Finally, we consider the universal relation for the $f$-mode frequency $\sigma_c$ observed in the rotating frame for sequences of constant baryon mass, given by Eq.~(5) in \cite{Kokkotas2020},
\begin{equation}
\label{eq:kk2}
    \frac{\sigma_c}{\sigma_0} = 1+b_1 \left(\frac{\Omega}{\Omega_{\rm max}}\right)+b_2\left(\frac{\Omega}{\Omega_{\rm max}}\right)^2,
\end{equation}
where $(b_1,b_2)=(0.517,-0.542)$ for the counterrotating branch and $(b_1,b_2)=(-0.235,-0.491)$ for the corotating branch. In contrast to \cite{Kokkotas2020}, we normalize the angular velocity $\Omega$ 
by its maximum rotation limit $\Omega_{\rm max}$ instead of the the Keplerian limit $\Omega_K$,
since $\Omega_{\rm max}$ can be larger than $\Omega_K$ by about 2\% for QSs as we have discussed. 
The ambiguity between the two values does not arise in \cite{Kokkotas2020} as 
$\Omega_{\rm max} = \Omega_K$ for NSs. 
Figure~\ref{fig:KK2ndUniversal} plots $\sigma_c /\sigma_0$ against $\Omega / \Omega_{\rm max}$ for 109 NS and QS models from various sequences of constant baryon mass. 
Let us recall that the mode frequencies observed in the rotating and inertial frames are related by 
$\sigma_c = \sigma_i + m \Omega / 2 \pi$. 
Contrary to Figs.~\ref{fig:KK3rdUniversal} and \ref{fig:KK1stUniversal}, the corotating modes are now represented by the lower branch of data in Fig.~\ref{fig:KK2ndUniversal}. 
Our SFHo NS data still satisfy Eq.~(\ref{eq:kk2}) very well, but the QS data deviate a lot from the fitting relations. 
However, it should be pointed out that the spread of the data around the upper gray line at high rotation rates is similar to that of the original NS data used in \cite{Kokkotas2020} to produce the fitting curve (see Fig. 2 in \cite{Kokkotas2020}). 

For the corotating modes (lower branch), the QS data deviate significantly from Eq.~(\ref{eq:kk2}). 
While realistic NS models generally cannot rotate fast enough to reach the onset of viscosity-driven instability, marked by the purple horizontal line where $\sigma_c=0$ in the figure,
Fig.~\ref{fig:KK2ndUniversal} shows that the QS data cross the purple line shortly before reaching the maximum rotation rate. 
In retrospect, the deviation between the NS and QS data at high rotation rates may be associated with the fact that there is an upper bound of the spin parameter $j \sim 0.7$ for realistic NSs when $\Omega \approx \Omega_{\rm max}$ \cite{Lo_2011}, while there is no such bound for QSs (see also Fig.~\ref{fig:jfk}). 
As Eq.~(\ref{eq:kk2}) was originally proposed to fit realistic NSs only \cite{Kokkotas2020}, 
the equation would not be able to cover QS models with $j \gtrsim 0.7$. We shall show below that 
a better universal relation for the corotating modes satisfied by NSs and QSs can be obtained by invoking the spin parameter directly. 


\subsubsection{Critical values of the spin parameter, energy ratio, and eccentricity}
\begin{figure}[t]
    \centering
    \includegraphics[width=\columnwidth]{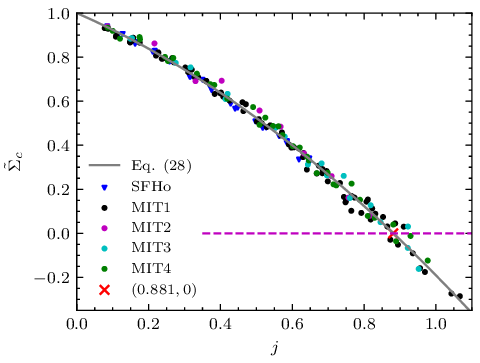}
    \caption{Normalized frequency $\tilde{\Sigma}_c$ [Eq.~(\ref{eq:sigmac})]
    is plotted against the spin parameter $j$ in the rotating frame.
    It contains both constant central density and constant baryon mass sequences, including 167 models as in Fig.~\ref{fig:KK3rdUniversal}.
    The quadratic fitting curve [Eq.~(\ref{eq:Sigma_c_j})] crosses the zero-frequency point at $j\approx 0.881$.}
    \label{fig:SJ}
\end{figure}

We now investigate further the onset of the viscosity-driven instability for rotating QSs. 
As it is expected to be difficult for realistic NSs to rotate fast enough to achieve this instability before 
reaching the Keplerian limit, the onset of this instability is a special (if not unique) phenomenon for rapidly rotating QSs among stellar objects. 
To determine the onset of the instability, which is close to the maximal rotation rate where physical quantities become sensitive to the angular frequency, we first propose a fitting relation which relates the corotating mode frequency $\sigma_c$ in the rotating frame to the spin parameter $j$ with relatively small variance.

We first define (in the code units) a scaled frequency for the corotating mode 
\begin{equation}
\label{eq:sigmac}
    \tilde{\Sigma}_c = \frac{2\pi {M}\sigma_c}{-0.0047+0.133\ \eta+0.575\ \eta^2} , 
\end{equation}
where $\eta=\sqrt{M^3/I}$ is the effective compactness originally introduced in \cite{Lau2010} for 
nonrotating stars, but here it is generalized to rotating stars. 
The denominator on the right-hand side of Eq.~(\ref{eq:sigmac}) is motivated by the universal relation between $\eta$ and the scaled $f$-mode angular frequency $2 \pi M \sigma_0$ for nonrotating NSs and QSs \cite{Lau2010}. Note that we have corrected a typographical error in the coefficient of $\eta^2$ in Eq.~(6) of \cite{Lau2010}.
In Fig.~\ref{fig:SJ}, we plot $\tilde{\Sigma}_c$ against the spin parameter $j$ for both constant 
central energy density and constant baryon mass sequences. 
Comparing to the corotating modes plotted in Fig.~\ref{fig:KK2ndUniversal}, the NS and QS data are now ``unified" and can be fit by
\begin{equation}
    \tilde{\Sigma}_c = -0.477 j^2 -0.714 j + 1 .
    \label{eq:Sigma_c_j}
\end{equation}
The root-mean-square of the residuals is $0.0251$. 
The rapidly rotating QS data with $j \gtrsim 0.7$ behave as if they are merely an extension of NS data to higher spin parameters. 
Our fitting curve crosses the zero-frequency point at $j\approx 0.881$, which represents the onset
of the viscosity-driven instability for both sequences of constant central energy density and 
constant baryon mass. 

\begin{figure}[t]
    \centering
    \includegraphics[width=\columnwidth]{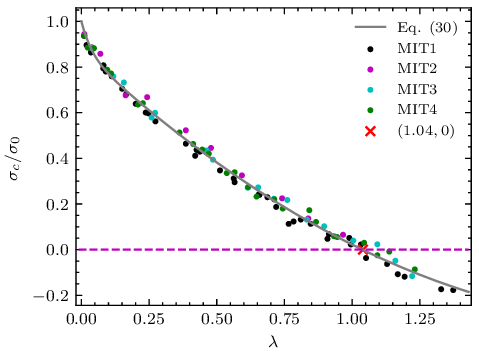}
    \caption{Plot of the scaled corotating mode frequency $\sigma_c/\sigma_0$ in the rotating frame
    against the normalized energy ratio $\lambda = (T/|W|)/(T/|W|)_{\rm crit}$ for constant baryon mass sequences. The data set contains 94 QS models used in Fig.~\ref{fig:KK2ndUniversal}.
    The fitting curve [Eq.~(\ref{eq:TWfitting})] crosses the zero-frequency line at $\lambda \approx 1.04$.}
    \label{fig:STW}
\end{figure}

Traditionally, the onset of the instability is characterized by the critical value of the ratio between the rotational kinetic energy and gravitational potential energy $T/|W|$ and the eccentricity 
$\zeta = (1-(r_{\rm p}/r_{\rm eq})^2)^{1/2}$, where $r_{\rm p}$ and $r_{\rm eq}$ are the polar and equatorial 
coordinate radii, respectively.   
As discussed in Sec.~\ref{sec:intro}, the Newtonian limit $(T/|W|)_{\rm crit,Newt} = 0.1375$
was obtained for Maclaurin sequences \cite{CHANDRASEKHAR_1969}, while general relativity weakens the instability by 
increasing the critical energy ratio \cite{PhysRevD.66.044021}.
An approximate relation for the critical energy ratio was obtained in~\cite{PhysRevD.66.044021} for constant baryon mass sequences of homogeneous incompressible bodies in general relativity,
\begin{equation}  
\left({T}/{|W|}\right)_{\rm crit} = \left({T}/{|W|}\right)_{\rm crit,Newt} + 0.126\ \chi\left(1+\chi\right),
\label{eq:TW_crit}
\end{equation} 
where $\chi=M/R$, $M$, and $R$ are the compactness, gravitational mass, and radius of the corresponding 
nonrotating model, respectively.
The difference between the relativistic and Newtonian critical values of $T / |W|$ is about 20\% for compactness $\chi \approx 0.2$.

QSs described by the MIT bag model can be approximated very well by homogeneous incompressible bodies \cite{Gondek2003}.
To check whether our QS data can also be approximated by Eq.~(\ref{eq:TW_crit}), we plot the scaled corotating mode frequency $\sigma_c / \sigma_0$ in the rotating frame against the
normalized energy ratio $\lambda \equiv (T/|W|)/(T/|W|)_{\rm crit}$ for constant baryon mass 
sequences in Fig.~\ref{fig:STW}. 
The trend of the numerical data can be fitted by 
\begin{equation}
\label{eq:TWfitting}
    \frac{\sigma_c}{\sigma_0}=1 + 0.130 (e^{-27.3 \lambda} - 1) - 1.10 \lambda + 0.256 \lambda^2. 
\end{equation}
The root-mean-square of the residuals is $0.0233$.
In addition to a quadratic fitting, an exponential function is included to take into account the fast initial decrease. The fitting curve crosses the zero point at $\lambda \approx 1.04$, meaning that the critical value
for our QS models is only 4\% higher than the approximate value of homogeneous incompressible bodies
predicted by Eq.~(\ref{eq:TW_crit}).

\begin{figure}[t]
    \centering
    \includegraphics[width=\columnwidth]{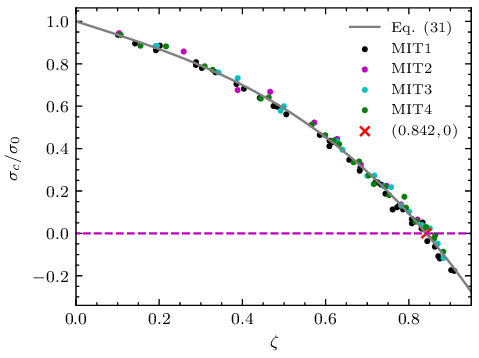}
    \caption{Plot of the scaled corotating mode frequency $\sigma_c/\sigma_0$ in the rotating frame
    against the eccentricity $\zeta$ for constant baryon mass sequences. The data set contains 94 QS models used in Fig.~\ref{fig:KK2ndUniversal}.
    The fitting curve [Eq.~(\ref{eq:eccfitting})] crosses the zero-frequency line at 
    $\zeta \approx 0.842$.}
    \label{fig:ecc}
\end{figure}

On the other hand, the critical value of eccentricity depends weakly on the compactness, it should thus be close to the Newtonian critical value $\zeta_{\rm crit,Newt}=0.8127$~\cite{PhysRevD.66.044021,Gondek2003}. 
In Fig.~\ref{fig:ecc}, we plot the scaled mode frequency $\sigma_c/\sigma_0$ against the eccentricity $\zeta$.
Its fitting curve is
\begin{eqnarray}
    \label{eq:eccfitting}
   \frac{\sigma_c}{\sigma_0}  = 1 - 0.622 \zeta - 0.799\zeta^3,
\end{eqnarray}
with the root-mean-square of the residuals being 0.0207. The fit predicts the onset of the instability at 
$\zeta\approx 0.842$, which is about $3.6\%$ higher than the Newtonian value. 

\subsection{Fitting relations of $p$-modes}
\label{sec:pmode}

We end this section by also providing fitting relations for the first $p$-modes of rotating QSs which were strongly excited in our simulations. 
As it is well known that $p$-modes are more EOS sensitive~\cite{Andersson_1998} and dependent strongly on the density and pressure profiles, universal relations are not expected to exist for them. 
Since our generalized MIT bag model EOS contains only two parameters, namely the 
bag constant $B$ and the square of the speed of sound $c_{ss}$, it is possible to find fitting 
relations for the $p$-modes by invoking these parameters. 
Furthermore, it is found that dimensionless frequencies like $Mp$ (with $p$ being the $p$-mode frequency) are independent of the bag constant in the MIT bag model~\cite{chan2013thesis} as it is just a scaling factor, and hence only $c_{ss}$ will be relevant to our fitting relations.

\begin{figure}[t]
    \centering
    \includegraphics[width=\columnwidth]{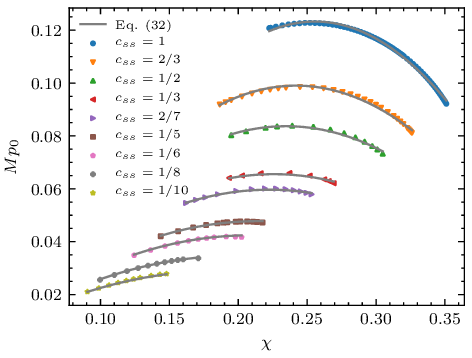}
    \caption{Plot of scaled frequency $M p_0$ of nonrotating models against compactness $\chi=M/R$ for the class of MIT bag EOSs with the square of the speed of sound $c_{ss}$ ranging from $1/10$ to $1$. The gray fitting curves are based on Eq.~(\ref{eq:p0fit}). }
    \label{fig:p0UR}
\end{figure}

This can be illustrated by considering the $p$-mode frequency $p_0$ of nonrotating QSs. We found that the scaled frequency $M p_0$ of our nonrotating QS models can be fitted well by 
\begin{eqnarray}
\label{eq:p0fit}
   M p_0 &=& ( a_1 c_{ss} + a_2) \chi^2  + (a_3 c_{ss}^2 + a_4 c_{ss} + a_5     \cr \cr 
   && +\ a_6 /c_{ss})\chi  +  a_7 c_{ss}^2 ,
\end{eqnarray}
where $M$ and $\chi = M / R$ are the gravitational mass and compactness, respectively. 
The seven fitting parameters are $a_1=-2.700$, $a_2=-0.5845$, $a_3=0.2183$, $a_4=1.202$, $a_5=0.2664$, $a_6=-0.006893$ and $a_7=-0.09141$. This relation is obtained by fitting to nonrotating QS data with
$M \geq 1.4 M_\odot$ and different values of $c_{ss}$ ranging from $1/10$ to 1 as shown in Fig.~\ref{fig:p0UR}. The root-mean-square of the residuals is $0.000450$.

\begin{figure}[t]
    \centering
    \includegraphics[width=\columnwidth]{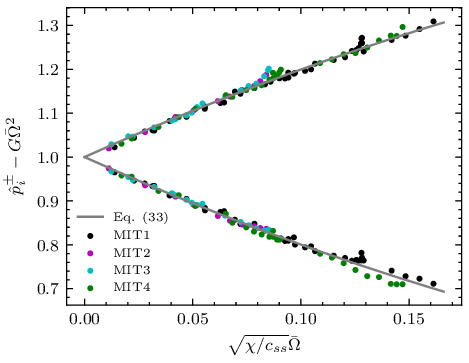}
    \caption{Plot of $\hat{p}_i^{\pm} - G \bar{\Omega}^2 $ against 
    $\sqrt{{\chi}/{c_{ss}}}\bar{\Omega}$ for constant baryon mass sequences of rotating QSs. 
    The fitting relations [Eq.~(\ref{eq:puniversal})] for the $m=\pm 2$ $p$-mode
    frequencies $p^{\pm}_i$ observed in the inertial frame are given by the solid lines.   }
    \label{fig:onerulesthemall}
\end{figure}

For the $l=2$ $p$-modes of rotating QSs, we use the following ansatz for the 
$m= 2$ ($m=-2$) $p$-mode frequencies $p^{+}_i$ ($p^{-}_i$) observed in the inertial frame for constant baryon mass sequences:
\begin{equation}
\label{eq:puniversal}
  \hat{p}_i^{\pm} = 1\mp  F(\chi,c_{ss},\bar{\Omega}) \bar{\Omega}+ G(\chi,c_{ss}) \bar{\Omega}^2 ,
\end{equation} 
where $\hat{p}_i^{\pm}=p_i^{\pm}/{p_0}$, $\bar{\Omega}=\Omega/(2\pi p_0)$, and $p_0$ is the 
$p$-mode frequency of the corresponding nonrotating star with gravitational mass $M$ and 
compactness $\chi$.
The two functions $F(\chi,c_{ss},\bar{\Omega})$ and $G(\chi,c_{ss})$ are given by
\begin{equation}
    F(\chi,c_{ss},\bar{\Omega}) = b_1 \sqrt{\frac{\chi}{c_{ss}}} +  b_2 {\frac{\chi}{c_{ss}}}\bar{\Omega} ,
\end{equation}
   and
\begin{equation}
   G(\chi,c_{ss}) = (b_3 c_{ss}^2 + b_4 )\chi + (b_5 c_{ss}^2 + b_6)  ,
\end{equation}
where the fitting parameters are $b_1=2.22$, $b_2=-2.24$, $b_3=82.1$, $b_4=44.6$, $b_5=-28.8$, and $b_6=-11.4$. 
To illustrate the fitting relation, we plot $\hat{p}_i^{\pm} - G \bar{\Omega}^2 $ against 
$\sqrt{{\chi}/{c_{ss}}}\bar{\Omega}$ in Fig.~\ref{fig:onerulesthemall}. The numerical data can 
be fitted well by Eq.~(\ref{eq:puniversal}) with the root-mean-square of residues being $0.00741$ for the upper branch and $0.00701$ for the lower branch.
It should be noted that the above fitting parameters are obtained by excluding those rapidly rotating degenerate models close to the maximum rotation limit illustrated in Fig.~\ref{fig:jfk}.

In reality, it is not expected to be able to detect the $p$-modes of compact stars from their emitted
GW signals anytime soon, even with the next generation of detectors. However, it might still be interesting
to consider how one could (in principle) make use of these fitting relations.
As an illustration, let us first ignore the rotational effects and assume that the $f$-mode frequency 
(and its damping time) and the $p$-mode frequency of a nonrotating compact star are observed. 
Applying the universal relations for the $f$-mode of nonrotating stars in \cite{Lau2010}, which are 
valid for both NSs and QSs, the mass $M$ and radius $R$, and hence the compactness $\chi$ 
of the star, can then be inferred approximately. Equation~(\ref{eq:p0fit}) can then be solved for the single variable $c_{ss}$ and one can check whether the observation data is consistent to our generalized MIT bag model. For instance, an inferred value of $c_{ss}=1/3$ would mean that the star is consistent 
with a QS described by the canonical MIT bag model. 
On the other hand, an inferred value of $c_{ss}$, which is far away from our fitting range of Eq.~(\ref{eq:p0fit}), would serve as strong evidence against our QS models.

Similarly, if the angular velocity $\Omega$ and the two frequencies $p^{\pm}_i$ are observed for 
a rotating compact star, then Eq.~(\ref{eq:puniversal}) can be used to relate the three parameters
$p_0$, $c_{ss}$, and $\chi$. If the star is slowly rotating so that its compactness can be 
approximated by the value of the nonrotating counterpart, then one can solve
for $p_0$ and $c_{ss}$. We can then compare the inferred value of $p_0$ to the observed frequency of
the $m=0$ axisymmetric $p$-mode (if it is available), which is well approximated by $p_0$ for 
a slowly rotating star, and determine whether the observed compact star is consistent with our QS models.


\section{Conclusion}
\label{sec:conclude}

The sharp high-density surface of a bare QS presents a great challenge for grid-based hydrodynamical 
modeling of the star. In this paper, building on top of the numerical relativity code 
\texttt{Einstein Tookit}, we have implemented a numerical method based on a positivity-preserving Riemann solver and a dustlike EOS for the atmosphere to perform stable evolutions of rapidly rotating QSs
in general relativity. Our work represents a new addition to the list of just a few 
fully general relativistic simulations of QSs available up to today \cite{Zhenyu2021,Enping2021,Enping2022}. 
   
The fidelity of our method has been tested and confirmed by comparing the oscillation mode frequencies of nonrotating QSs extracted from simulations with the results obtained from perturbative calculations. 
The $f$-mode of rapidly rotating QSs are investigated in details. In particular, we find that two of 
the universal relations for the $l=|m|=2$ nonaxisymmetric modes proposed originally for rotating NSs
\cite{Kokkotas2020} are still valid for QSs (see Figs.~\ref{fig:KK3rdUniversal} and \ref{fig:KK1stUniversal}). However, the QS data deviate significantly from another universal relation for the corotating modes observed in the rotating frame (see Fig.~\ref{fig:KK2ndUniversal}). 
In addition to the $f$-modes, we have also studied the first $p$-modes of rotating QSs. For QSs
described by our generalized MIT bag model, we report fitting relations for the $p$-mode 
frequencies of both nonrotating and rotating stars. 

We also find that, when considering sequences of constant central energy density, the onset of the CFS instability for QSs occurs when the angular velocity $\Omega \approx 3.4 \sigma_0$, which agrees
with the finding for NSs \cite{Kokkotas2020}. 
In addition to the CFS instability, we have also studied the viscosity-driven instability of QSs. 
We find that the onset of the instability for rotating QSs occurs when the spin parameter $j \approx 0.881$ for both sequences of constant central energy density and constant baryon mass. 
For QS sequences of constant baryon mass , we also find that the critical value of the ratio between 
the rotational kinetic energy and gravitational potential energy $T/|W|$ for the onset of the 
instability agrees with the value predicted for homogeneous incompressible bodies in general 
relativity to within 4\%, 
and the critical value of the eccentricity $\zeta$ is only $3.6\%$ larger than the Newtonian value~\cite{PhysRevD.66.044021}. 
Realistic NSs are generally not expected to be able to rotate fast enough to trigger this instability before 
reaching the Keplerian limit. This can be seen from Fig.~\ref{fig:KK2ndUniversal} that the NS
data for the frequencies of the corotating modes $\sigma_c$ observed in the rotating frame do not 
cross zero before the Keplerian limit. 
The universal relation between the spin parameter and $\tilde{\Sigma}_c$, which is just a rescaled 
$\sigma_c$, proposed by us in Eq.~(\ref{eq:Sigma_c_j}) can unify the NS and QS data and also
predict the onset of the instability to occur at $j \approx 0.881$ as shown in Fig.~\ref{fig:SJ}. 
The fact that realistic NSs cannot trigger the instability can be associated to the existence 
of an upper bound $j\sim 0.7$ for uniformly rotating NSs \cite{Lo_2011}.

\begin{acknowledgments}
We thank Hoi-Ka Hui for useful discussions and Shu-Yan Lau for sharing his oscillation code for 
us to compute the mode frequencies of nonrotating stars for benchmarking. 
This work is partially supported by a grant from the Research Grants Council of the Hong Kong
Special Administrative Region (Project No. 14304322).
We also acknowledge the support of the CUHK Central High Performance Computing Cluster, on which our 
simulations were carried out.  
\end{acknowledgments}

%
\end{document}